\documentclass[a4paper]{quantumarticle}

\pdfoutput=1
\usepackage[utf8]{inputenc}
\usepackage[english]{babel}
\usepackage{tikz}
\usepackage{lipsum}
\usepackage{amsmath, amssymb}
\usepackage{braket}
\usepackage{graphicx}
\usepackage{xcolor}
\usepackage[linesnumbered,ruled,vlined]{algorithm2e}
\usepackage{hyperref}

\begin{document}

\title{Research of the Variational Shadow Quantum Circuit Based on the Whale Optimization Algorithm in Image Classification}
\author{Shuang Wu}
\affiliation{College of Computer and Information Science, Chongqing Normal University,Chongqing, 401331, China}
\orcid{0009-0000-5701-9210}
\author{Xueliang Song}
\affiliation{College of Computer and Information Science, Chongqing Normal University,Chongqing, 401331, China}
\author{Yumin Dong}
\email{dym@cqnu.edu.cn}
\orcid{0000-0002-7890-4427}
\affiliation{College of Computer and Information Science, Chongqing Normal University,Chongqing, 401331, China}
\author{Fanghua Jia}
\affiliation{Qingdao University of Technology, Shandong, 266033, China}
\maketitle
\onecolumn

\begin{abstract}
In order to explore the possibility of cross-fertilization between quantum computing and neural networks as well as to improve the classification performance of quantum neural networks, this paper proposes an improved Variable Split Shadow Quantum Circuit (VSQC-WOA) model based on the Whale Optimization Algorithm. In this model, we design a strongly entangled local shadow circuit to achieve efficient characterization of global features through local shadow feature extraction and a sliding mechanism, which provides a rich quantum feature representation for the classification task. The gradient is then computed by the parameter-shifting method, and finally the features processed by the shadow circuit are passed to the classical fully connected neural network (FCNN) for processing and classification. The model also introduces the Whale Optimization Algorithm (WOA) to further optimize the weights and biases of the fully connected neural network, which improves the expressive power and classification accuracy of the model. In this paper, we firstly use different localized shadow circuit VSQC models to achieve the binary classification task on the MNIST dataset, and our design of strongly entangled shadow circuits performs the best in terms of classification accuracy. The VSQC-WOA model is then used to multi-classify the MNIST dataset (three classifications as an example), and the effectiveness of the proposed VSQC-WOA model as well as the robustness and generalization ability of the model are verified through various comparison experiments. 
\end{abstract}

\section{Introduction}
\label{sec_one}

With the arrival of the big data era and the rapid development of artificial intelligence technology, deep learning [1] has demonstrated powerful capabilities in many fields. However, with the increase of data size and computational complexity, traditional computational methods gradually show limitations and are difficult to cope with the demands of massive data processing and high-complexity problems. Quantum computing, as a new computing paradigm that utilizes the principles of quantum mechanics, has attracted the attention of a large number of researchers, driving the development of quantum machine learning (QML), an emerging cross-cutting field [2-4].

Quantum machine learning combines the properties of quantum parallelism, quantum entanglement, and quantum superposition with the powerful modelling capabilities of classical machine learning to provide novel solutions to machine learning tasks. However, existing quantum machine learning models still face many challenges in processing complex data, especially in feature extraction and representation. Traditional quantum feature extraction methods are difficult to fully mine the potential structural information of the data, resulting in limited model classification performance. Therefore, how to effectively combine the advantages of quantum feature extraction and classical machine learning methods to construct efficient quantum-classical hybrid models has become an important direction of current research [5-7].

In recent years, experts and scholars have invested a great deal of effort in studying a variety of quantum neural network models based on medium-scale quantum processors [8-10]. Such as the quantum perceptual machine model [11], the quantum tensor neural network [12], etc. These quantum neural network models simulate classical quantum systems with network structure characteristics in quantum Hilbert space, build different quantum circuits to approximate nonlinear functions, and realize arbitrary nonlinear quantum neurons based on a generalizable model framework [13]. Due to the wide application of quantum neural networks in the field of machine learning. In 2018, Farhi et al. [14] constructed an early quantum neural network model based on quantum gates, demonstrating its feasibility on recent quantum computers while exploring the implementation of classification tasks and their potential. Grant et al. [15], on the other hand, proposed a hierarchical quantum neural network classifier model based on a multiscale entanglement reformulation of the Ansatz. They used this model to perform several binary classification tasks on the MNIST dataset, including classification of 0s and 1s, classification of 2s and 7s, determination of whether a number is greater than 4, and parity determination. These experiments show that this hierarchical quantum neural network achieves good performance in handling different binary classification tasks. In 2019, Cong et al. [16], inspired by classical convolutional neural networks (CNNs), designed a quantum convolutional neural network (QCNN), which uses parameterized quantum circuits to construct the network layer with modules such as quantum convolution, quantum pooling, and quantum all-connectivity, which significantly improves the ability of quantum models to represent features. In 2020, YaoChong Li et al. [17] combined parameterized quantum circuits with convolutional neural networks to propose a hybrid quantum-classical trained recognition model, which achieved good experimental results. Henderson et al. [18] took another approach; they used random quantum circuits to replace the convolutional layers in traditional convolutional neural networks (CNNs) to construct a quantum hybrid structure. This structure utilizes quantum computing properties such as superposition and entanglement for more efficient feature extraction when performing input feature extraction on handwritten digits. 2021, several researchers have further advanced the theory and application of quantum neural networks. Kashif et al. [19] proposed to embed variational quantum circuits into a classical neural network as a hidden layer, thus designing a hybrid quantum-classical neural network; Pesah et al. [20] theoretically analyzed the nature of parameter gradient scaling in QCNNs, proved the trainability of the model, and avoided the barren plateau problem; Abbas et al. [21] verified the superiority and training efficiency of quantum neural networks on real-world quantum devices by introducing the concept of effective dimensionality and relating the Fisher information spectrum of a quantum neural network to the model performance. In 2023, Qu Z et al. [22] combined quantum blockchain technology with quantum convolutional neural networks to develop a quantum arrhythmia detection system (QADS), which achieved efficient detection of abnormal heartbeats and demonstrated robustness and accuracy. In 2024, researchers continued to explore new applications and optimizations of quantum neural networks. For example, Song Z et al. [23] converted two-dimensional tensor networks (TNs) into quantum circuits for supervised learning and constructed tensor network-inspired quantum circuits (TNQCs), which performed well on the corresponding datasets. Park S et al. [24] designed AQUA, an analytics-driven method for software validation of quantum neural networks, and verified its stability and interpretability in the field of autonomous driving. Fan F et al. [25] optimized a hybrid quantum-classical convolutional neural network (QC-CNN) by introducing a magnitude coding technique, demonstrating superior classification performance. Wu Q et al. [26] constructed a QCNN with enhanced circuits, which effectively improves the global feature extraction capability by designing a global quantum convolutional kernel and a decreasingly parameterized quantum pooling layer. In addition, Shi M et al. [27] optimized a hybrid quantum neural network model in a multi-class image classification task, which performed outstandingly in migration learning experiments on IBM quantum hardware, although the performance was slightly lower than the best classical CNN. In summary, the superposition and entanglement properties of quantum computing are widely used to alleviate the overfitting problem caused by insufficient data, which provides a solid theoretical foundation and a broad application prospect for quantum machine learning.

With the rapid development of quantum neural networks, it is gradually becoming possible to design quantum classification models that can solve complex classification tasks. However, many key issues still need to be further investigated, such as how to construct more general quantum circuit models, how to deal with diverse data types, how to optimize the hyperparameters to improve the training effect of the models, how to improve the simulability of the quantum models with the classical models and how to implement more effective optimization algorithms to alleviate the impact of the barren plateau phenomenon on the trainability of the models, the breakthroughs in these research directions will not only further promote the practicality of quantum neural networks, but also lay the foundation for the widespread application of quantum computing in machine learning. further promote the practicalization of quantum neural networks, and will also lay the foundation for the wide application of quantum computing in the field of machine learning.

The main innovations of this paper are as follows:

Innovation 1: By designing a strongly entangled quantum state evolution circuit, an efficient feature extraction function has been realized, and the sliding property of the quantum circuit further extends its adaptability so that it can flexibly deal with data distribution.

Innovation 2: Optimize the parameters of the classical post-processing network by combining it with the Whale Optimization Algorithm (WOA). The algorithm simulates the feeding behaviour of beluga whales and searches for the optimal solution through mechanisms such as random swimming and spiral bubble net attack, which improves the parameter tuning efficiency of the network, thus significantly improving the performance of the model.

Innovation 3: The whale optimization algorithm was compared with a variety of optimization algorithms (including particle swarm algorithm, genetic algorithm, artificial immunity algorithm, etc.) for experiments. The results show that the WOA algorithm performs optimally in terms of convergence speed and classification accuracy, verifying its superiority and practical value in image classification tasks.

The structure of the paper consists of the following sections. Section 1 is to introduce the background, purpose, and significance of the research of the paper. Section 2 describes the theoretical knowledge related to the research and provides support for the subsequent content. Section 3 introduces the structure and implementation steps of the VSQC-WOA model. Section 4 empirically investigates the MNIST dataset using the VSQC-WOA model in order to verify the validity of the model. Section 5 summarizes the main research findings of the paper and provides suggestions for future research.

\section{Introduction to relevant theories}
\label{sec_two}

\subsection{Classical neural network}

Artificial neural network (ANN), or classical neural network (NN) [28-29], is a computational model inspired by biological neural networks. It consists of a large number of “neurons” or nodes that interact with each other through a complex network of connections. Each connection carries a weight indicating the efficiency of information transfer or the strength of the connection, thus modelling the working mechanism of biological neural networks.

The workflow of a classical neural network consists of several stages, including data preprocessing, feedforward propagation, backpropagation, model training, and testing and evaluation. Firstly, data preprocessing is a key step, which usually normalizes and denoises the data and divides it into training and testing sets to improve the training efficiency and generalization ability of the model. Next, the data is fed into the network through feed-forward propagation and processed sequentially through neurons in the input, hidden, and output layers. The neurons in the hidden layer weight and sum the data passed from the previous layer and then complete the nonlinear transformation through the activation function until the output layer generates the final prediction. Subsequently, it enters the backpropagation phase, where the weights of the model are adjusted by calculating the error (loss) between the predicted value and the target value. The error will propagate backward from the output layer, layer by layer, and the weights are gradually optimized according to the gradient of the error. In order to accelerate the learning process, optimization algorithms such as Gradient Descent SGD, Adam, or RMSprop are usually used, which ultimately lead to a gradual reduction of the error and convergence of the model.

After training, classical neural networks are usually evaluated for model performance using test sets. Commonly used metrics include accuracy, recall, and F1 score, which are used to measure the model's ability to generalize to new data as well as the effectiveness of the classification task. Classical neural networks are especially suitable for small-scale datasets or scenarios with limited resources due to their simple structure and easy implementation and are widely used in image recognition, speech recognition, and natural language processing. Compared to traditional machine learning methods, ANN is able to learn complex nonlinear patterns, reducing the reliance on manual feature engineering while significantly accelerating the training process with the help of optimization algorithms. This makes ANN very suitable for deployment in systems that require real-time response and becomes an efficient tool for solving complex data problems. The classical neural network structure is shown in Figure 1.

\begin{figure}
\centerline{\includegraphics[width=320pt,height=17pc]{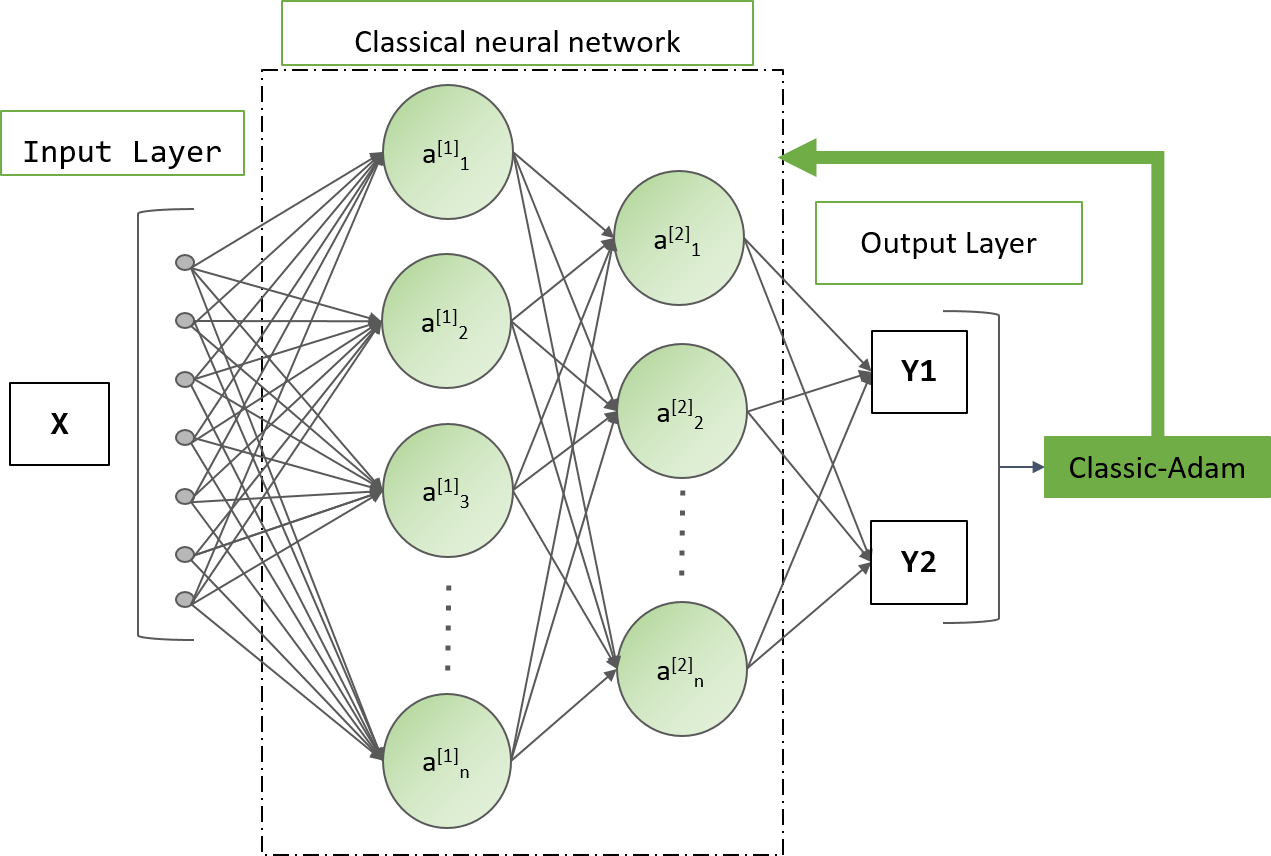}}
\caption{\footnotesize The structure of a classical neural network consists of three parts. On the left side is the input layer, which is used to receive external data or feature vectors X;in the middle is the hidden layer, which contains multiple neurons $a_n^{[1,2]}$ that are connected to each other as well as to the input layer by weights; and on the right side is the output layer, which consists of two neurons,$\mathrm{Y1}$ and $\mathrm{Y2}$, which correspond to the two different output categories or predicted values. The training of the whole network is performed by means of the classical optimization algorithm SGD, where the goal of the optimization is to constantly update the parameters in order to minimize the loss function. \label{figure1}}
\end{figure}

\subsection{Quantum neural network}
Quantum neural network (QNN) [30-31] combines quantum computing and classical machine learning by parameterizing quantum circuits (PQC) for training and optimization. It uses qubits and quantum gates to construct models and optimizes performance by tuning the parameters of these quantum gates. QNN uses a variational approach to find the optimal parameters by minimizing the loss function, allowing quantum circuits to better represent the data. It typically employs a hybrid classical-quantum architecture, using classical computing resources for gradient computation and optimization and quantum computing resources to process the data and compute the expectation value of the quantum state. Similar to classical neural networks, QNN uses a loss function to measure the difference between the predicted results and the true results and uses classical optimization algorithms to adjust the parameters of the quantum circuits. QNN is widely used in the fields of quantum classification, quantum regression, and quantum reinforcement learning, and is particularly suitable for dealing with high-dimensional data and complex pattern recognition problems. The structure of the quantum neural network is shown in Figure 2.

\begin{figure}
\centerline{\includegraphics[width=200pt,height=13pc]{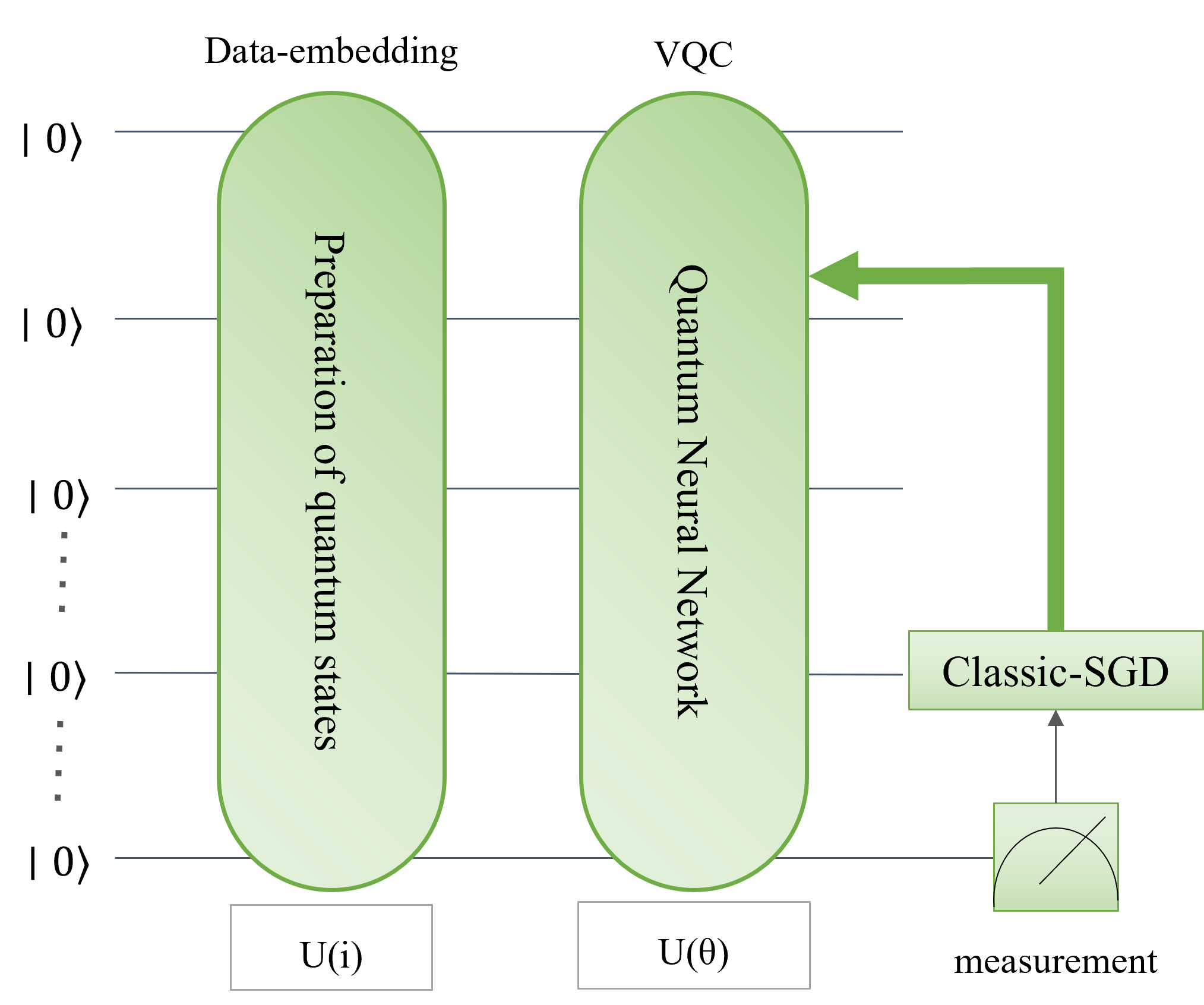}}
\caption{\footnotesize Quantum Neural Network.In this structure, classical data is first prepared into quantum states using angle encoding, and then a quantum circuit (ansatz) containing multiple parameters is processed. Finally, the measurement structure is input into the optimizer of a classical computer to update the parameters in the quantum circuit before calculation, until the set optimization termination condition is reached. $U_{i}$ represents a quantum state preparation circuit. This is a quantum circuit used to prepare an initial quantum state (usually a $|0\rangle$ state) as an input state. This section is mainly responsible for encoding input data into quantum states to prepare for subsequent quantum computing. $U_{\theta}$ stands for quantum neural network circuit. This is a quantum circuit used to perform quantum computing and process input states. It contains a series of quantum gates and parameterized rotation gates, used to simulate weights in neural networks and perform quantum calculations to achieve a specific computing task or training process. The number of layers in $U_{\theta}$ is greater than or equal to 1. \label{figure2}}
\end{figure}

\subsection{Quantum computing}

The basic unit of information in quantum computing [32] is the qubit, whose state can be manipulated by quantum logic gates. qubits have superposition and entanglement properties, which form the core foundation of the powerful computational capabilities of quantum computers. Information in quantum computing can be represented by n qubits on space $C^{2^n\times2^n}$, mathematically described by a semipositive definite matrix $\rho\succeq0$ with property $\mathrm{Tr}(\rho)=1$. A quantum state is pure if the rank of this density matrix is $\mathrm{Rank}(\rho)=1$; otherwise it is a mixed state. For a pure state $\rho$, it can be expressed in terms of a unit vector in the sense of $\rho=|\psi\rangle\langle\psi|$ that where the ket notation $|\psi\rangle\in C^{d}$   denotes a column vector and bra notation $\langle\psi|=|\psi\rangle^{\dagger}$ with $\dagger$  denoting conjugate transpose. For simplicity of representation, pure states are also often directly represented by $\left|\psi\right\rangle$ .Mixed states, on the other hand, can be represented by $\rho=\sum_{i}q_{i}|\psi_{i}\rangle\langle\psi_{i}|$ . The coefficients $q_{i}\geq0$ record the probability   $|\psi_{i}\rangle\langle\psi_{i}|$ that the quantum system is in each corresponding pure state, and hence the mixed state is also $\sum_iq_i=1$.

A quantum gate [33] is a fundamental unit of operation in quantum computing that changes the state of a qubit by acting on the quantum state through a linear transformation. Unlike logic gates in classical computing, quantum gates can manipulate superposition states (i.e., linear combinations of qubits in both 0 and 1 states) as well as entanglement states (non-classical correlations between qubits) of quantum states. They are usually represented as matrices to describe the operations applied to a quantum state. The quantum gates mainly used in this paper include quantum rotating X-gate, quantum rotating Y-gate, quantum rotating Z-gate and quantum controlling non-gate, which play a key role in transforming quantum states. These quantum gates cover operations from single-bit to multi-bit, where quantum non-gates and quantum rotating gates are used for single-bit state tuning, while quantum controlling non-gates are used for multi-bit inter-controls, which are the core tools of quantum computing.

Quantum NOT gates, commonly known as X-gates or Pauli-X gates, are used to flip the state of qubits. The Pauli matrix  $\sigma_x=\begin{pmatrix}0&1\\1&0\end{pmatrix}$.Let $|\varphi\rangle=\binom{cos\theta_0}{sin\theta_0}$,the quantum NOT gate $\sigma_x$ acts to $|\varphi\rangle$ as: $\sigma_x|\varphi\rangle=\begin{pmatrix}0&1\\1&0\end{pmatrix}\begin{pmatrix}cos\theta_0\\sin\theta_0\end{pmatrix}=\begin{pmatrix}sin\theta_0\\cos\theta_0\end{pmatrix}$.

Quantum rotation gates are used to perform rotation operations on qubits in quantum computing. These gates change the quantum state of qubits by applying rotation to them. Rotating gates are commonly used to achieve precise control of quantum states in quantum circuits, forming the methods of Rx, Ry, and Rz.

The Rx-gate, Ry-gate and Rz-gate are generated by the Pauli-X, Pauli-Y and Pauli-Z matrices as generating elements which have the following matrix form:
\begin{equation}R_x(\theta)=\mathrm{e}^{\frac{-\mathrm{i}\theta X}{2}}=\cos{(\frac{\theta}{2})}\mathrm{l}-\mathrm{i}\sin{(\frac{\theta}{2})}X=
\begin{pmatrix}
\cos{(\frac{\theta}{2})} & -\mathrm{i}\sin{(\frac{\theta}{2})} \\
-\mathrm{i}\sin{(\frac{\theta}{2})} & \cos{(\frac{\theta}{2})}
\end{pmatrix}\end{equation}
\begin{equation}R_y(\theta)=\mathrm{e}^{\frac{-i\theta Y}{2}}=\cos{(\frac{\theta}{2})}\mathrm{I}-\sin{(\frac{\theta}{2})}Y=
\begin{pmatrix}
\cos{(\frac{\theta}{2})} & -\sin{\frac{\theta}{2}} \\
\sin{(\frac{\theta}{2})} & \cos{(\frac{\theta}{2})}
\end{pmatrix}\end{equation}
\begin{equation}R_z(\theta)=\mathrm{e}^{\frac{-\mathrm{i}\theta Z}{2}}=\cos\left(\frac{\theta}{2}\right)\mathrm{I}-\mathrm{i}\sin\left(\frac{\theta}{2}\right)\mathrm{Z}=
\begin{pmatrix}
\mathrm{e}^{\frac{-\mathrm{i}\theta}{2}} & 0 \\
0 & \mathrm{e}^{\frac{\mathrm{i}\theta}{2}}
\end{pmatrix}\end{equation}

A CNOT gate is used to implement conditional operations between qubits. It consists of a control bit and a target bit. The CNOT gate flips the state of the target bit when the control bit is in state $|1\rangle$; if the control bit is in state $|0\rangle$, the target bit remains unchanged. the matrix of the CNOT gate is denoted as:
\begin{equation}\mathrm{CNOT}=
\begin{pmatrix}
1 & 0 & 0 & 0 \\
0 & 1 & 0 & 0 \\
0 & 0 & 0 & 1 \\
0 & 0 & 1 & 0
\end{pmatrix}\end{equation}

\section{Application of the Improved VSQC Model in Image Classification}
\label{sec_three}
\subsection{Overall structure of the VSQC-WOA model}

The variational shadow quantum circuit [34-36] is a hybrid model structure that combines quantum computing and classical machine learning, and its core idea is to use the properties of the shadow circuit to extract the quantum shadow features of the data, and to perform subsequent processing and classification using classical neural networks. The model designed in this paper is based on the Whale Optimization Algorithm (WOA) [37-38], which further optimizes the parameters of the classical post-processing network to improve the overall performance of the model.

The model consists of two parts: a quantum feature extractor and a classical post-processor. The quantum feature extractor is implemented based on multilayer variational quantum circuits (VQCs). Firstly, the data is mapped into quantum states by preprocessing, and then local shadow features are extracted by a specific variational quantum circuit $U(\boldsymbol{\theta})$ (denoted as shadow circuit). The circuit design contains multiple rotating gates (Rx, Ry, Rz) and control non gates (CNOT), and the position of the circuit action is dynamically adjusted through a sliding window mechanism to gradually scan the entire quantum state for localized features. The local shadow features are obtained by measuring the expectation values associated with predefined observable measurements (e.g., Pauli operators).

In the optimization process, the parameter shift method is used to calculate the gradient for each trainable parameter in the shadow circuit. By applying small shifts (forward and reverse) on the circuit parameters, the difference in the expected values is computed to estimate the gradient. This gradient information helps in tuning the parameters in the shadow circuit to improve the classification performance.

The classical post-processor employs a shallow neural network (FCNN) with a fully connected layer to map shadow features extracted from quantum circuits to classification labels. To optimize the parameters of this part, this paper introduces the whale optimization algorithm, which simulates the whale's hunting behaviour and dynamically adjusts the neural network weights and biases to minimize the loss function. Specifically, the WOA effectively avoids local optimality by guiding the whale population towards the optimal position, updating the individual positions, and reducing the global loss. The design of the model not only shows the potential of quantum computing, but also provides new ideas for optimizing the classical quantum hybrid model. The structure of the variational shadow quantum circuit model based on the whale optimization algorithm is shown in Figure 3.

\begin{figure}
\centerline{\includegraphics[width=500pt,height=20pc]{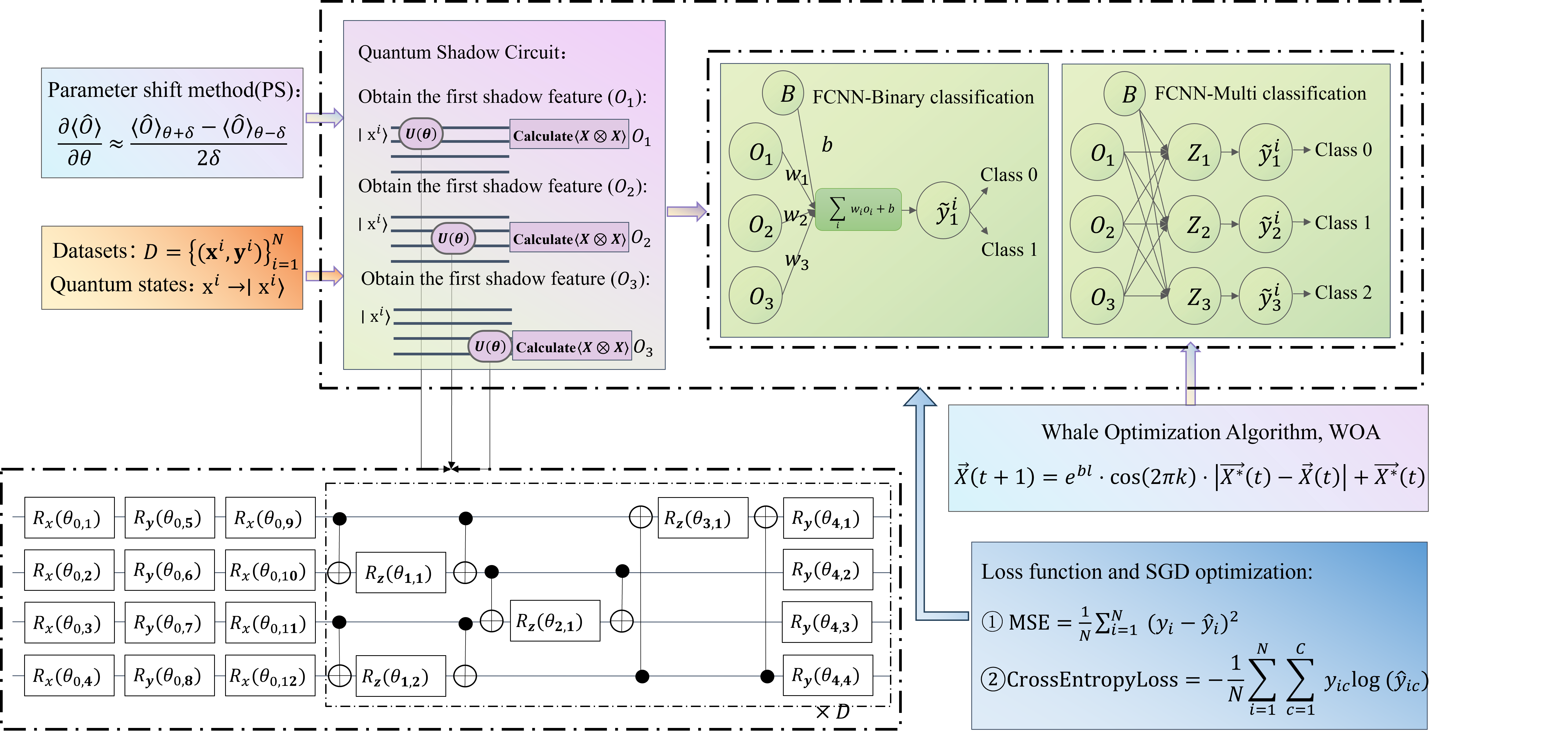}}
\caption{\footnotesize Structure of the VSQC-WOA based model. Firstly, the input dataset $D=\{(x_i,y_i)\}_{i=1}^n$ is encoded into a quantum state by a quantum circuit $x_i\to|x_i\rangle$ . The quantum shadow circuit generates multiple shadow features {$O_1,O_2,O_3$} and passes them to the fully connected neural network (FCNN). In the classification task, the binary classification network is responsible for binary classification and outputs the category {0,1}, and the multiclassification network outputs the multiclass {0,1,2,...}.The parameters in FCNN are optimized by the Whale Optimization Algorithm (WOA), which optimizes the loss function by minimizing the loss function (Mean Square Error MSE or CrossEntropy Loss CrossEntropyLoss) combined with the parameter shift method and stochastic gradient descent (SGD) to complete the optimization. The structure of the quantum circuit is designed with strong entanglement to generate efficient quantum state evolution through a series of rotating and controlled gates.\label{figure3}}
\end{figure}

\subsection{Implementation of Binary Classification Based on Improved VSQC Model}

\subsubsection{Data preprocessing}

Since each handwritten digit image consists of 28×28 gray scale pixel points and each pixel value is in the range of [0,255]. These two-dimensional images are first converted to one-dimensional vectors x. The pixel values of each image are sequentially arranged into a one-dimensional array $\mathbf{x}=[x_1,x_2,\ldots,x_d]$  by a flattening operation. To ensure that all image vectors have the same length, we add zeros to the end of the spread vector to bring it to the desired fixed length.
 
Next, the spread image vector is normalized. We achieve normalization by calculating its L2 norm and dividing each element of the vector by this norm. This step adjusts the length of the image vectors to 1 and avoids the effect of data magnitude differences on model training. We start with L2-paradigm normalization to obtain the unit vector:
\begin{equation}\mathbf{x}_{\mathrm{norm}}=\frac{\mathbf{x}}{\|\mathbf{x}\|_{2}}\end{equation}

Where $\|\mathbf{x}\|_2$ is the L2 paradigm of the vector $\mathrm{x}$, calculated as:
\begin{equation}\parallel\mathrm{x}\parallel_2=\sqrt{\sum_iv_i^2}\end{equation}

After completing the data normalization, we use amplitude coding to convert the classical data into quantum states. The process of amplitude coding maps each element of the classical data to the amplitude of the quantum state, thus providing input data for subsequent quantum computing tasks.

The classical data vector $\mathbf{x}= \begin{bmatrix} x_1,x_2,...,x_d \end{bmatrix}$ is encoded as a quantum state $\left|\psi\right\rangle$ whose amplitude corresponds to the components of the data vector x. The amplitude of the quantum state is encoded as $\left|\psi\right\rangle$. The amplitude encoded quantum state can be expressed as: 
\begin{equation}\mid\psi\rangle=\frac{1}{\sqrt{N}}\sum_{i=0}^{N-1}x_i\mid i\rangle\end{equation}

where $\frac{1}{\sqrt{N}}$ is the normalization factor that ensures that the total probability amplitude of the quantum state is 1 (unit amplitude). For image data, the data vector $x_\mathrm{norm}$ will be mapped to the amplitude of the quantum state:   
\begin{equation}\mid\psi\rangle=\sum_ix_{\mathrm{norm},i}\mid i\rangle\end{equation}

\subsubsection{\label{sec:levelb}Improved VSQC model}
\textbf {VSQC modelling process}. We now demonstrate the structure of the VSQC model used for the binary classification task, which is shown in Figure 4. In the training process of VSQC, firstly, the input is a training dataset $\mathcal{D}^{(train)}:=\{(\rho_{in}^{(m)},y^{(m)})\}_{m=1}^{N_{train}}$  represented by n qubits, where $y^{(m)}\in\{0,1\}$ denotes that each data point is encoded as a density matrix $\rho_{in}^{(m)}$ with corresponding binary labels $y^{(m)}$,
Then, a local shadow circuit $n_{qsc}$ acts on the previous $n_{qsc}$ qubits and estimates the bubble ley expectation $\operatorname{Pauli-}(X\otimes\cdots\otimes X)$ as the first shadow feature $O_{1}$. Next, the same shadowing circuit is implemented on the subspace from $2^{nd}$ the $(2+n_{qsc}-1)^{th}$ qubits to extract the second shadowing feature $O_{2}$.
When the shadowing circuit slides down, we get a total of $n-n_{qsc}+1$ shadowing features. This convolutional operation of sliding through qubit positions can be flexibly adapted.In addition, although one shadow circuit is used by default, it can be adapted to more complex classification tasks by increasing the number of shadow circuits $(n_{s})$ with $n_{s}(n-n_{qsc}+1)$ shadow features. The extracted shadow features $\{O_{i}\}$ are subsequently fed into the classical FCNN for processing. In FCNN, these characteristics are weighted and summed by weights $w\in R^{n-n_{qsc}+1}$ and biases $b\in R$ and then assigned to a fixed range $\hat{y}^{(m)}\in[0,1]$ by a sigmoid activation function $\sigma(z)=\left(1+\mathrm{e}^{-z}\right)^{-1}$. For each input data, the cumulative loss between the predicted value $\hat{y}^{(m)}$ and its true label $y^{(m)}$ is calculated  $\mathcal{L}(\boldsymbol{\theta},\boldsymbol{w},b;\mathcal{D}^{(train)})$. Subsequently, the gradient-based optimization algorithm SGD is used to update the parameters $\theta$ of the shadow circuit and the parameters w and b of the neural network simultaneously to minimize the loss function step by step. The VSQC completes the training process by repeating the above training steps until the loss function converges to the tolerance $\Delta\mathcal{L}\leq\varepsilon$ or other stopping conditions are satisfied. The VSQC training pseudocode is shown in Algorithm 1.

\begin{figure}
\centerline{\includegraphics[width=250pt,height=8pc]{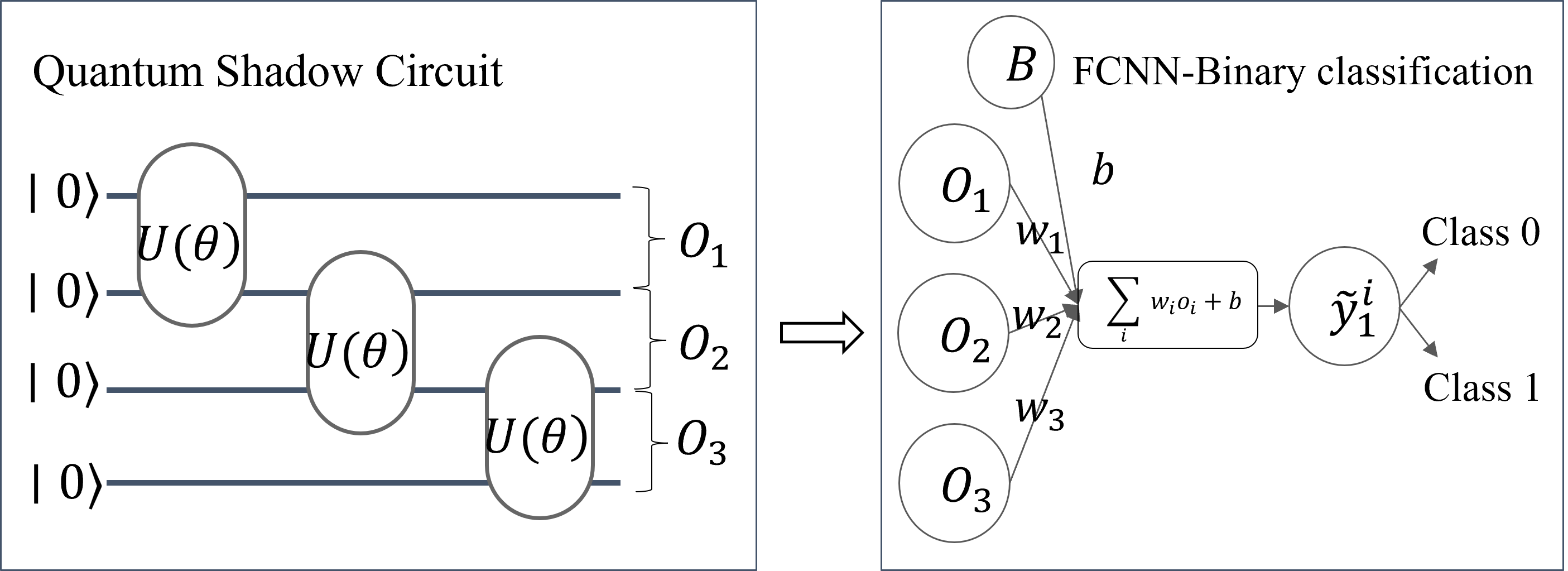}}
\caption{\footnotesize Structural diagram of the variational shadow quantum circuit (VSQC) for binary classification of $n=4,n_{qsc}=2$ and K = 3. The details are as follows: In a quantum device, a shadow circuit acts on the subspace of the input state $\rho_{in}$ , performs measurements and extracts shadow features. Throughout the system, the shadow circuit slides over the different subspaces of the Hilbert space and collects the expectation value of input state $\mathrm{Pauli-}(X\otimes X)$ , the generated “shadow features”. In the classical device, the generated shadow feature $o_{i}$$\mathrm{'s}$ is fed into a fully connected neural network for processing. In a binary classification task, the output $\hat{y}$ of the neural network is a value between 0 and 1, which is used for classification judgment. Note that the parameters and design of all sliding shadow circuits $U(\theta)$$\mathrm{'s}$ in a Hilbert space of n qubits are identical, thus ensuring consistent and reproducible results.\label{figure4}}
\end{figure}

\begin{algorithm}[ht]
\caption{VSQC for Binary Classification: The Training Process}
\label{alg:VSQC}
\KwIn{Training dataset $\mathcal{D}^{(\text{train})} = \{(\rho_{\text{in}}^{(m)}, y^{(m)}) \in [0, 1]\}_{m=1}^{N_{\text{train}}}$; \\
\hspace{1em} number of qubits $N$, shadow circuit depth $D$; \\
\hspace{1em} number of shadow qubits $n_{\text{qsc}}$, learning rate $LR$; \\
\hspace{1em} number of epochs $EPOCH$, batch size $BATCH$}
\KwOut{Trained parameters $\theta$, $w$, and $b$}

Initialize parameters $\theta$, $w$, and $b$ randomly\;
\For{$ep \leftarrow 1$ \KwTo $EPOCH$}{
    Shuffle the training dataset $\mathcal{D}^{(\text{train})}$\;
    \For{$itr \leftarrow 1$ \KwTo $N_{\text{train}}$ with step size $BATCH$}{
        Select a batch $\{(\rho_{\text{in}}^{(m)}, y^{(m)})\}_{m=itr}^{itr+BATCH-1}$\;
        \For{$m \leftarrow itr$ \KwTo $itr + BATCH - 1$}{
            Apply shadow circuit $U(\theta)$ to $\rho_{\text{in}}^{(m)}$\;
            Measure expectations $\langle X \otimes X \rangle$, record as $o_i^{(m)}$\;
            Feed $o_i^{(m)}$ into classical NN, get $\hat{y}^{(m)} \in [0,1]$\;
            Compute loss $L(\hat{y}^{(m)}, y^{(m)})$\;
        }
        Compute average loss over the batch\;
        Update $\theta$, $w$, and $b$ using gradient descent\;
    }
    Evaluate training and test accuracy on $D^{(\text{test})}$\;
}
\Return{$\theta$, $w$, and $b$}\;
\end{algorithm}

\textbf{Strongly entangled VSQC}. In this paper, we design a strongly entangled localized shadow circuit $U_{\theta}$, as shown in Figure 5. The circuitry operates parametrically on the $n_{qsc}$ qubit through a series of quantum gates with the goal of extracting quantum features and achieving entanglement-enhanced state evolution. The circuit first applies parameterized single-qubit rotational gates $\mathrm{R_x}(\theta)=e^{-i\theta x/2},\quad\mathrm{R_y}(\theta)=e^{-i\theta y/2}$
to each active qubit, which act to introduce parameterized local rotational operations and adjust the amplitude and phase of the quantum state. Subsequently, the circuit performs a series of CNOT gates (e.g., $\mathrm{CNOT}(q_0,q_1)$     and ring connection $\mathrm{CNOT}(q_3,q_0))$, local rotation $\mathrm{R_z}(\phi)=e^{-i\phi z/2}$ for further tuning of the phase of the states in conjunction with the final $R_y$ gates to enhance the expressive power, which create strong correlations between the qubits to ensure global entanglement between the qubits and enhance the expressive power of the circuit. Acting on the initial state $|\psi_{\mathrm{in}}\rangle$ yields the output state $|\psi_{\mathrm{out}}\rangle=U_\theta|\psi_{\mathrm{in}}\rangle$.This structure can extract highly nonlinear and entangled features through parameter optimization, providing an efficient feature representation for quantum classifiers. The You evolution matrix of the whole circuit can be denoted as $U_\theta=\prod_{d=1}^\mathrm{depth}U_\mathrm{CNOT}U_\mathrm{R_z}U_\mathrm{R_x,R_y}$.For ease of differentiation, we name the circuit diagram Circuit-5.   
\begin{figure}
\centerline{\includegraphics[width=380pt,height=9pc]{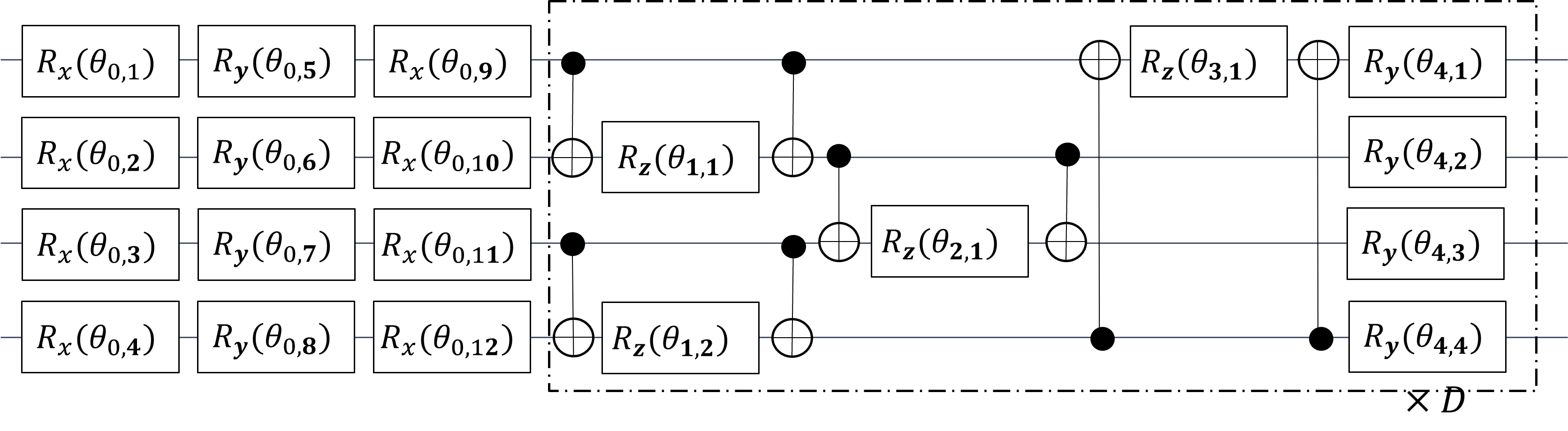}}
\caption{\footnotesize Structure of the strongly entangled localized shadow  Circuit-5. The first part of this circuit realizes a general rotational operation on the subspace of each single qubit by means of Rx-Ry-Rx combinations. The subsequent repeating module consists of a CNOT entanglement layer with Rz gates and a single qubit Ry rotation. To enhance the expressiveness of the quantum circuit, the modular circuit in the dashed box is repeated D times to form a multilayer structure, which significantly improves the expressiveness and adaptability of the circuit.\label{figure5}}
\end{figure}

In order to more fully verify the effectiveness of the localized shadow circuits we have designed, we have systematically compared them with several other mainstream quantum circuit architectures. The other circuits are shown in Figure 6. By evaluating the performance of these circuits under the same dataset and experimental conditions, we are able to gain a clearer understanding of the strengths and weaknesses of different circuit architectures in terms of feature extraction, entanglement generation, and model prediction accuracy. This comparison not only helps to quantify the performance of our circuits in specific tasks, but also provides a clear direction for further improvement. In particular, we analyze the effects of these factors on the model performance by adjusting the experimental parameters (e.g., circuit depth, number of parameters, etc.) to comprehensively verify the applicability and robustness of our designed circuits in various application scenarios.

\begin{figure}
\centerline{\includegraphics[width=470pt,height=18pc]{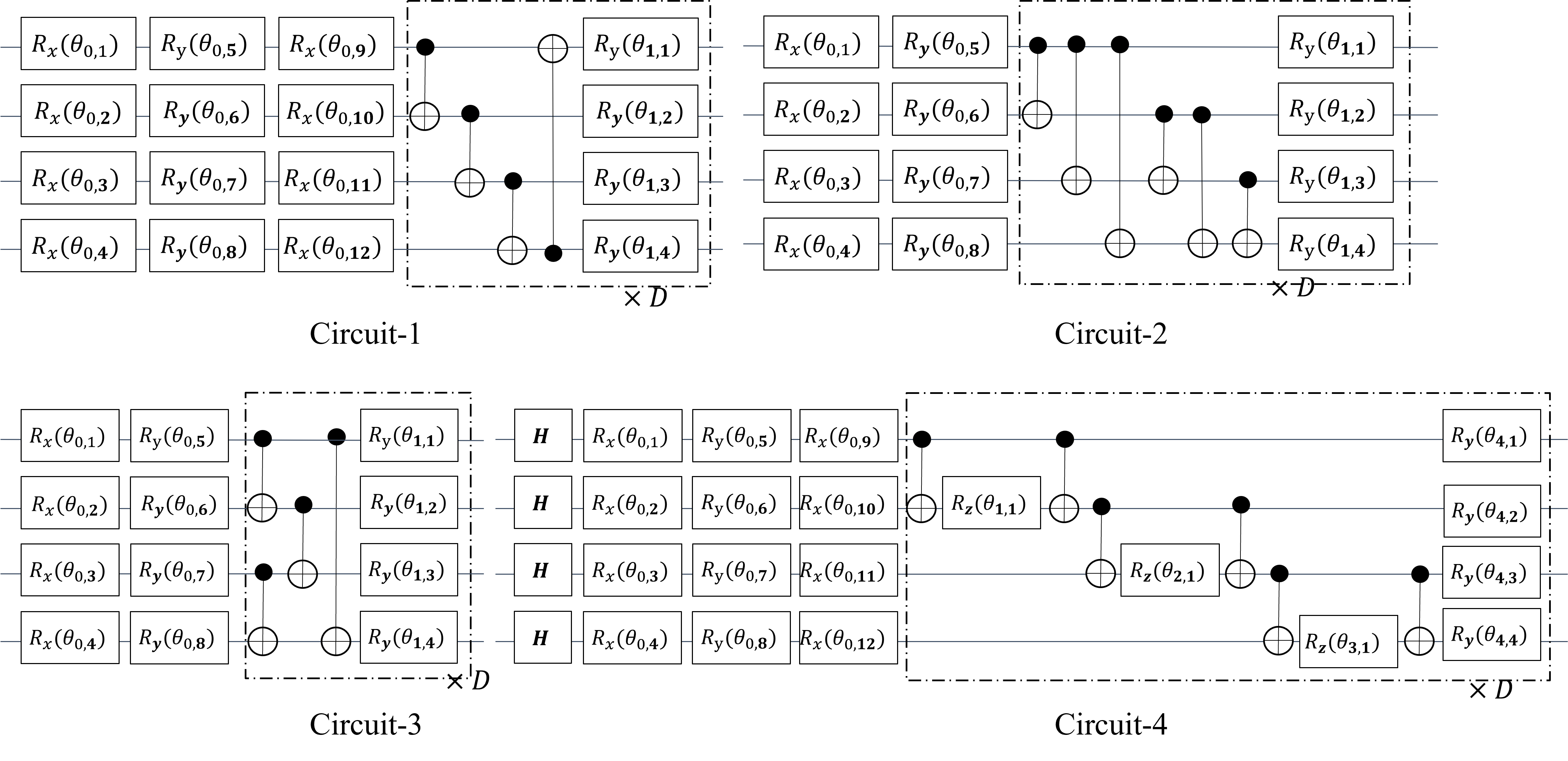}}
\caption{\footnotesize Structure of other shadow circuits.The first part of Circuit-1 [34] uses Rx-Ry-Rx combinations to represent general rotations on the subspace of each single qubit. The next repeating block consists of CNOT gates and single qubit Ry rotations.The first part of both Circuit-2 and Circuit-3 [39] uses Rx-Ry combinations to represent general rotations on the subspace of each single qubit. The next repeating block also consists of CNOT gates and single qubit Ry rotations.The first part of Circuit-4 [21] uses H-Rx-Ry-Rx combinations to represent general rotations on each single qubit subspace. The next repeating block consists of a CNOT entanglement layer with RZ gates and a single qubit Ry rotation. The block circuit in the dashed box is repeated D times to extend the expressive power of the quantum circuit.\label{figure6}}
\end{figure}

\textbf{Loss function.} In this paper, the loss function we use for binary classification is the mean square error loss function. 
\begin{equation}\mathcal{L}(\boldsymbol{\theta},\boldsymbol{w},b;\mathcal{D}):=\frac{1}{2N}\sum_{m=1}^{N}\left[\hat{y}^{(m)}\left(\rho_{in}^{(m)};\boldsymbol{\theta},\boldsymbol{w},b\right)-y^{(m)}\right]^{2}\end{equation}

where the predicted label $\hat{y}^{(m)}$ is defined as:\begin{equation}\hat{y}^{(m)}\left(\rho_{in}^{(m)};\boldsymbol{\theta},\boldsymbol{w},b\right):=\sigma\left(\sum_iw_io_i^{(m)}\left(\rho_{in}^{(m)};\boldsymbol{\theta}\right)+b\right)\end{equation}

$\sigma(z)$denotes the sigmoid activation function, and the shadow feature $o_{i}$ is computed: 
\begin{equation}o_{i}^{(m)}\left(\rho_{in}^{(m)};\boldsymbol{\theta}\right)=\mathrm{Tr}\left(\rho_{in}^{(m)}(\mathbb{I}\otimes\cdots\otimes U^{\dagger}(\boldsymbol{\theta})OU(\boldsymbol{\theta})\otimes\cdots\otimes\mathbb{I})\right)\end{equation}

It is important to note that the shadow circuit $\mathcal{U}(\theta)$ and the physical observables $O=X\otimes\cdots\otimes X$ always act on the same local qubits to ensure that the extracted quantum features are consistent with the local properties of the target state. Specifically, the shadow circuit $\mathcal{U}(\theta)$ can usually be further decomposed into a series of parameterized $U(\theta)$ operators,
\begin{equation}U(\theta)=\prod_{l=L}^1U_l(\theta_l)V_l.\end{equation}

 Where $U_l(\theta_l)=\exp(-i\theta_lP_l/2)$,$V_{l}$ denote fixed operators, such as CNOT gates, etc. $P_{l}$
denote the three kinds of revolving gates, X,Y,Z.

This decomposition not only conforms to the physical constraints of current quantum hardware, but also facilitates the implementation of gradient optimization for efficient tuning of the circuit parameters and accurate evaluation of the expectation value of local quantum states. This design ensures flexibility in feature extraction while enhancing the circuit's ability to express complex quantum states.Optimization Algorithm. In this paper, the optimization algorithm we use is the gradient descent based optimization algorithm SGD  [40-41], for each input $\rho_{in}^{(m)}$
\begin{equation}\frac{\partial\mathcal{L}}{\partial w_i}=\left(\hat{y}^{(m)}-y^{(m)}\right)\cdot\hat{y}^{(m)}\left(1-\hat{y}^{(m)}\right)\cdot o_i^{(m)},\end{equation}
\begin{equation}\frac{\partial\mathcal{L}}{\partial b}=\left(\hat{y}^{(m)}-y^{(m)}\right)\cdot\hat{y}^{(m)}\left(1-\hat{y}^{(m)}\right), \end{equation}
\begin{equation}\frac{\partial\mathcal{L}}{\partial\theta_{l}}=\frac{\partial\mathcal{L}}{\partial\hat{y}^{(m)}}\cdot\sum_{i}\frac{\partial\hat{y}^{(m)}}{\partial o_{i}^{(m)}}\cdot\frac{\partial o_{i}^{(m)}}{\partial\theta_{l}}=\left(\hat{y}^{(m)}-y^{(m)}\right)\cdot\sum_{i}\hat{y}^{(m)}\left(1-\hat{y}^{(m)}\right)w_{i}\cdot\frac{\partial o_{i}^{(m)}}{\partial\theta_{l}}
\end{equation}

In the above formulation, the partial derivatives of weight $w_{i}$ and bias b are explicitly defined, and they will be directly derived by gradient computation in a classical computing device. The values of these partial derivatives will be used as key elements in the optimization process to quantify the sensitivity of the current parameters to the loss function. Specifically, using the backpropagation algorithm, we can efficiently propagate the gradient of the loss function from the output layer back to the input layer, calculating and accumulating the gradient of each weight $w_{i}$ and bias b layer by layer. The parameters are updated in each iteration to progressively minimize the target loss function. Through such a process, the classical device is able to efficiently optimize the neural network parameters, working in concert with the parameter optimization of the quantum circuit to achieve continuous improvement in overall model performance.

\subsection{\label{sec:levelb}Implementing Multivariate Classification Based on the VSQC-WOA Model}

In this section, we will briefly introduce the VSQC-WOA framework for multi-label classification tasks. Since the data preprocessing process in the multivariate classification task is basically the same as that of binary classification, and the overall structure and working mechanism of the Variable Scale Quantum Shadow Circuit (VSQC) continues the design of binary classification, we will not repeat these same parts in this section. We will focus on key details and improvements in the multivariate classification scenario that differ from binary classification. Firstly, we will show the overall structural picture of VSQC-WOA when dealing with multi-label tasks through a structural diagram, followed by a detailed discussion of the loss function design applicable to multi-labeling, focusing on how to deal with outputs with multiple labels. Next, we will analyze the parameter shift method for computing the gradient and the whale algorithm for optimizing the weights and biases in classical fully connected. Based on this, we will ensure efficient model training and parameter updating in complex multi-label classification tasks. With these extensions and improvements, VSQC demonstrates its adaptability and strong performance in multi-label classification tasks.

\subsubsection{\label{sec:levelb}VSQC-WOA model}

In this subsection, we describe in detail the basic structure and working principle of the Variable Scale Quantum Shadow Circuit (VSQC) for the multi-label classification task, as shown in  Figure 7. Meanwhile, in order to present more intuitively the practical application of VSQC in multi-label classification tasks, we will also provide the pseudo-code of the algorithm, as shown in Algorithm 2, to fully demonstrate its corresponding training process. From the whole process of data preprocessing to model training to parameter updating, the algorithm will clearly illustrate the details and computational logic of each step.

\begin{figure}
\centerline{\includegraphics[width=250pt,height=8pc]{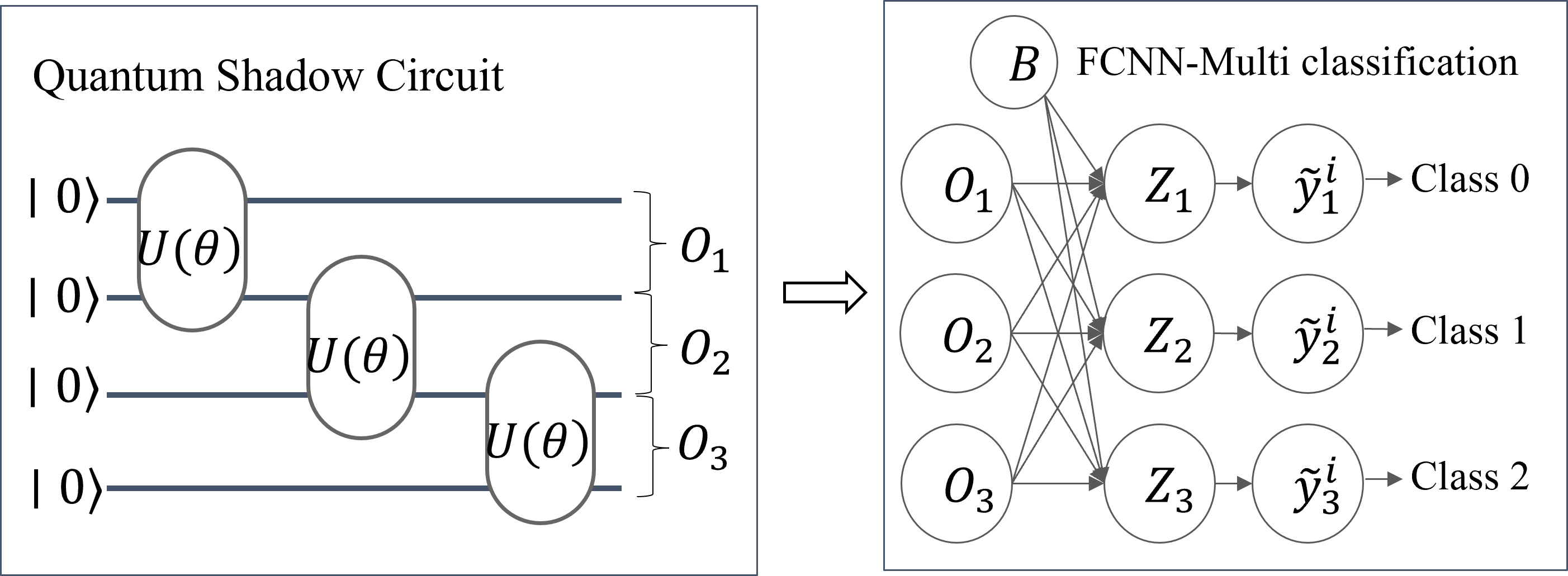}}
\caption{\footnotesize Architecture of the $n=4,n_{qsc}=2$ and the variational shadow quantum circuit (VSQC) for multi-label classification with K = 3. In the quantum device, the shadow circuit acts on the subspace of the input state $\rho_{in}$  to realize the variational quantum operation. By moving the shadow circuit throughout the system, the expectation value of input state $\mathrm{Pauli-}(X\otimes X)$,i.e., the generated “shadow feature”, can be collected. These shadow features $O_{i}$ are then passed to a fully connected neural network (FCNN) in a classical device. In FCNN, the features are mapped to the output of multilabel classification using a softmax activation function, where the output $\hat{y}$ is a K-dimensional vector representing the predicted results of multilabel classification.\label{figure7}}
\end{figure}

\begin{algorithm}[ht]
\caption{VSQC-WOA for Multi-Classification: The Training Process}
\label{alg:VSQC-WOA}
\KwIn{Training dataset $\mathcal{D}^{(\text{train})} = \{(p_{\text{in}}^{(m)}, y^{(m)}) \in \{0, 1, 2\}\}_{m=1}^{N_{\text{train}}}$; \\
\hspace{1em} test dataset $\mathcal{D}^{(\text{test})}$, validation dataset $\mathcal{D}^{(\text{val})}$; \\
\hspace{1em} number of qubits $n$, circuit depth $D$; \\
\hspace{1em} number of whales $num\_whales$, maximum iterations $max\_iters$}
\KwOut{Trained parameters $\theta$, $w$, and $b$}

\BlankLine
\textbf{Initialization}\;
Initialize quantum circuit parameters $\theta$ and classical NN parameters $w$, $b$\;
Construct variational quantum circuit $U(\theta)$ with $n$ qubits and depth $D$\;

\BlankLine
\textbf{Shadow Feature Extraction}\;
\For{$m \leftarrow 1$ \KwTo $N_{\text{train}}$}{
    Encode $p_{\text{in}}^{(m)}$ into quantum state $\rho_{\text{in}}^{(m)}$\;
    Apply $U(\theta)$ to $\rho_{\text{in}}^{(m)}$\;
    Measure expectations $\langle X \otimes X \rangle$, record as shadow features $o_i'$\;
}

\BlankLine
\textbf{Classical Neural Network Training}\;
Feed shadow features $o_i'$ into fully connected neural network\;
Train using cross-entropy loss to optimize $w$ and $b$\;
Update $\theta$ using gradient descent\;

\BlankLine
\textbf{Whale Optimization Algorithm}\;
Initialize population of whales (positions and velocities)\;
\For{$t \leftarrow 1$ \KwTo $max\_iters$}{
    \For{each whale $i \leftarrow 1$ \KwTo $num\_whales$}{
        Compute gradients for FCNN parameters using parameter shift method\;
        Calculate fitness (loss) for whale's position\;
        Update position based on attraction mechanism\;
    }
    Update best position if better solution found\;
}

\BlankLine
\textbf{Inference Process}\;
Compute test accuracy, precision, recall, and F1 score on $\mathcal{D}^{(\text{test})}$\;

\BlankLine
\Return{$\theta$, $w$, and $b$}\;
\end{algorithm}

\textbf{Loss function and gradient update. }Set up a given dataset $\mathcal{D}:=\{(\rho_{in}^{(m)},y^{(m)})\}_{m=1}^{N}\subset\mathbb{C}^{2^{n}\times2^{n}}\times\mathbb{R}^{K}$ and corresponding $n_{qsc}$ locally shadow circuits, where $y^{(m)}$
is a one-hot vector (one-hot vector) to represent the category to which the $m^{\mathrm{th}}$ data sample $\rho_{in}^{(m)}$ belongs. For example, with a total number of categories K = 3, $y^{(m)}=[1,0,0]^{\top}$ if the sample belongs to category 0,$y^{(m)}=[0,1,0]^{\top}$ if the sample belongs to category 1, and $y^{(m)}=[0,1,0]^{\top}$ for category 2. This representation clearly defines the category affiliation of each sample and provides the basis for the computation of the loss function. In the Variational Shadow Quantum Circuit (VSQC) framework for multi-label classification, the design of the loss function is based on the cross-entropy formulation. 
\begin{equation}\mathcal{L}(\boldsymbol{\theta},\boldsymbol{W},\boldsymbol{b};\mathcal{D}):=-\frac{1}{N}\sum_{m=1}^N\sum_{k=1}^Ky_k^{(m)}\log\hat{y}_k^{(m)}\left(\rho_{in}^{(m)};\boldsymbol{\theta},\boldsymbol{W},\boldsymbol{b}\right)\end{equation}  

Here, the output of the variational shadow quantum circuit (VSQC) is a K-dimensional vector 1 defined as follows:
\begin{equation}\hat{y}^{(m)}\left(\rho_{in}^{(m)};\boldsymbol{\theta},\boldsymbol{W},\boldsymbol{b}\right):=\sigma\left(\sum_{i=1}^{n-n_{qsc}+1}\boldsymbol{w}_{i}o_{i}^{(m)}\left(\rho_{in}^{(m)};\boldsymbol{\theta}\right)+\boldsymbol{b}\right)\end{equation} 

where $W=[\boldsymbol{w}_{1},\boldsymbol{w}_{2},\ldots,\boldsymbol{w}_{n-n_{qsc}+1}]\in\mathbb{R}^{K\times(n-n_{qsc}+1)}, \boldsymbol{b}\in\mathbb{R}^{K\times1}, \sigma(\boldsymbol{z})=\frac{\mathrm{e}^{\boldsymbol{z}}}{\sum_{j}\mathrm{e}^{z_{j}}}$ denotes the softmax activation function and the shadow feature $o_{i}$ is calculated as follows by: 
\begin{equation}o_{i}^{(m)}\left(\rho_{in}^{(m)};\boldsymbol{\theta}\right)=\mathrm{Tr}\left(\rho_{in}^{(m)}(\mathbb{I}\otimes\cdots\otimes U^{\dagger}(\boldsymbol{\theta})OU(\boldsymbol{\theta})\otimes\cdots\otimes\mathbb{I})\right)\end{equation}

for each data sample $(\rho_{in}^{(m)},y^{(m)})$in the dataset $\mathrm{D}$ and hypothesis $y_{k}^{(m)}=1$, the partial derivatives with respect to parameters $w_{ji},b_j$ and $\theta_{l}$ are calculated as follows:
\begin{equation}\frac{\partial\mathcal{L}(\boldsymbol{\theta},\boldsymbol{W},\boldsymbol{b};\rho_{in}^{(m)},y^{(m)})}{\partial w_{ji}}=
\begin{cases}
\left(\hat{y}_k^{(m)}-1\right)\cdot o_i^{(m)}, & j=k \\
\hat{y}_j^{(m)}\cdot o_i^{(m)}, & j\neq k 
\end{cases}\end{equation}
\begin{equation}\frac{\partial\mathcal{L}(\boldsymbol{\theta},\boldsymbol{W},\boldsymbol{b};\rho_{in}^{(m)},y^{(m)})}{\partial b_j}=
\begin{cases}
\left(\hat{y}_k^{(m)}-1\right), & j=k \\
\hat{y}_j^{(m)}, & j\neq k 
\end{cases}\end{equation}
\begin{equation}\frac{\partial\mathcal{L}(\boldsymbol{\theta},\boldsymbol{W},\boldsymbol{b};\rho_{in}^{(m)},y^{(m)})}{\partial\theta_{l}}=\sum_{i=1}^{n-n_{qsc}+1}\sum_{j=1}^{K}\left(\hat{y}_{j}^{(m)}w_{ji}-w_{ki}\right)\frac{\partial o_{i}^{(m)}\left(\boldsymbol{\theta};\rho_{in}^{(m)}\right)}{\partial\theta_{l}},\end{equation}

In the model, we employ the $\pi/2$ parameter shift rule [42], a method that enables accurate calculation of gradients on quantum devices. Specifically, the original parameter values of the circuit are first saved, and then small positive and negative offsets $\pm\delta$ are applied to the target parameters, respectively, to generate the corresponding quantum states. Next, the gradient of the parameters is estimated by calculating the expectation value of the specified observable measurements for the quantum states after forward and reverse shifts and using the difference between the two to estimate the gradient of the parameters. After the computation is completed, the circuit parameters are restored to the initial state to ensure the accuracy of subsequent computations. The rule converges faster and is more compatible with existing quantum hardware than traditional finite difference methods.

\subsubsection{\label{sec:levelb}Whale optimization algorithms}
The Whale Optimization Algorithm (WOA) is a population intelligence optimization algorithm based on the feed behaviourvior of humpback whales, inspired by their bubble net hunting strategy. In this paper, we introduce the Whale Optimization Algorithm (WOA), mainly for parameter optimization of classical post-processing networks. The algorithm first randomly initializes the population position, calculates the fitness value of each candidate solution, and records the best solution. In the search phase, by dynamically adjusting the control parameters A, C, and a, the whales randomly move to other positions in the population to expand the search range; when  $|\mathrm{~A~}|>1$, the algorithm enters the exploitation phase, where the whales update their positions by spiral paths around the current optimal solution or by linear contraction to mimic the hunting behaviour. After each iteration, the whale's fitness value is updated, and the new global optimal solution is recorded. The algorithm continues to iterate until it reaches the maximum number of iterations or the optimization objective and finally outputs the best solution. WOA is able to efficiently solve continuous or discrete complex optimization problems by balancing between global exploration and local exploitation, avoiding falling into local optimums, and improving the optimization efficiency and accuracy of the model.

The algorithm consists of two main phases; in the first phase (the development phase), surround prey and spiral update bits are implemented. In the second phase (exploration phase), finding the prey is done randomly. The mathematical model for each phase is shown below.

Surrounding prey. In the WOA algorithm, humpback whales perform a circling search around the prey after discovering its location. Since the location of the optimal solution in the search space is unknown, the algorithm assumes that the current optimal candidate solution is close to the target prey or optimal solution. As a result, other search agents (i.e., whales) try to adjust their position and move towards the current optimal solution with the aim of finding a better solution. This behaviour is modelled by a specific equation that simulates the process of a whale encircling its prey.
\begin{equation}\vec{X}(t+1)=\vec{X^*}(t)-\vec{A}\cdot\vec{D}\end{equation}
\begin{equation}\vec{D}=
\begin{vmatrix}
\vec{C}\cdot\vec{X^*}(t)-\vec{X}(t)
\end{vmatrix}\end{equation}

where $\overrightarrow{X^*}(t)$ denotes the whale's early optimal position at iteration t. $\vec{X}(t+1)$ is the whale's current position, and $\overrightarrow{D}$ is the distance vector between the whale and the prey, with $||$ denoting the absolute value. c and a are coefficient vectors, calculated as follows:
\begin{equation}\vec{A}=2\vec{a}\cdot\vec{r_{1}}-\vec{a}\end{equation}
\begin{equation}\vec{C}=2\cdot\vec{r_2}\end{equation}

where $\overrightarrow{a}$ is a linear decay factor, and $\vec{r_1}$ and $\vec{r_2}$ are random numbers obeying a uniform distribution of [0,1]. The value of  $\vec{A}$ can be in the interval (-a, a), where the value of a is reduced from 2 to 0 by iteration. $\vec{A}$ By choosing random values in the interval (-1, 1), the new position of any search agent can be determined somewhere between the current position of the best agent and the original position of the agent. Spiral update position. By calculating the distance between the whale position $(\mathrm{X},\mathrm{Y})$ and the prey position $(\mathbf{X}^*,\mathbf{Y}^*)$, the interval between them can be determined. Next, a spiral equation is generated between this positions to model the humpback whale's spiralling trajectory around its prey. This is shown below:
\begin{equation}\vec{X}(t+1)=e^{bl}\cdot\cos(2\pi k)\cdot\overrightarrow{D^{*}}+\overrightarrow{X^{*}}(t)\end{equation}
\begin{equation}\overrightarrow{D^*}=\left|\overrightarrow{X^*}(t)-\vec{X}(t)\right|\end{equation}

where b is a constant value used to identify the shape of the logarithmic spiral and l is a random number in the range [-1, 1]. This behaviour is represented in the WOA as changing the position of the whale during the optimization process. There is a 50\% chance of choosing between a shrink-wrap mechanism and a spiral model with their components designed as follows (where p is a random number in (0, 1)):
\begin{equation}\vec{X}(t+1)=
\begin{cases}
\vec{X^*}-\vec{A}\cdot\vec{D}, & \mathrm{if}p<0.5, \\
 \\
e^{bl}\cdot\cos{(2\pi k)}\cdot\vec{D^*}+\vec{X^*}(t), & \mathrm{if}p\geq0.5,  
\end{cases}\end{equation}

Finding prey. In searching for prey, whales find prey by randomizing their search based on each other's location, in a way that relies on the variance of the $\vec{A}$ vector. To avoid the search from falling into a local optimum, the WOA algorithm forces the search agent to move away from the current whale by using $\vec{A}$ vectors with random values grea ter or less than one. The entire exploration phase enhances the global search capability by realigning the positions of the search agents instead of relying only on the best search agent. This strategy helps the WOA algorithm to avoid local optima and enhances the efficiency of global search. The mathematical model is represented as follows:
\begin{equation}\vec{X}(t+1)=\overrightarrow{X_{\mathrm{rand}}}-\vec{A}\cdot\vec{D}\end{equation}
\begin{equation}\vec{D}=
\begin{vmatrix}
\vec{C}\cdot\vec{X_{\mathrm{rand}}}-\vec{X}
\end{vmatrix}\end{equation}

where $\overrightarrow{X_{\mathrm{rand}}}$ is a vector of random locations (random whales) chosen from the current population.

The pseudo-code of the whale optimization algorithm is shown in Algorithm 3:

\begin{algorithm}[ht]
\caption{Whale Optimization Algorithm (WOA)}
\label{alg:WOA}
\KwIn{Objective function $f(x)$; \\
\hspace{1em} Number of whales $N$; \\
\hspace{1em} Maximum iterations $Max\_iters$; \\
\hspace{1em} Search space bounds $[LB, UB]$}
\KwOut{Best solution $Best\_position$, Best fitness $Best\_score$}

\BlankLine
Initialize whale positions randomly within $[LB, UB]$\;
Initialize $Best\_position$ with a random position\;
Initialize $Best\_score = \infty$\;

\BlankLine
\For{$t \leftarrow 1$ \KwTo $Max\_iters$}{
    \For{$i \leftarrow 1$ \KwTo $N$}{
        Compute fitness $f(Position[i])$\;
        \If{$f(Position[i]) < Best\_score$}{
            $Best\_position \leftarrow Position[i]$\;
            $Best\_score \leftarrow f(Position[i])$\;
        }
        
        Compute parameters $A$, $C$, $p$\;
        $A \leftarrow 2a \cdot rand() - a$\;
        $C \leftarrow 2 \cdot rand()$\;
        $p \leftarrow rand()$\;
        
        \BlankLine
        \eIf{$p < 0.5$}{
            \If{$|A| < 1$}{
                $Position[i] \leftarrow Best\_position - A \cdot |C \cdot Best\_position - Position[i]|$\;
            }
            \Else{
                Select random whale $Position\_rand$\;
                $Position[i] \leftarrow Position\_rand - A \cdot |C \cdot Position\_rand - Position[i]|$\;
            }
        }{
            \textbf{Spiral motion:}\;
            $Distance\_to\_best \leftarrow |Best\_position - Position[i]|$\;
            $Position[i] \leftarrow Distance\_to\_best \cdot e^{b \cdot l} \cdot \cos(2 \pi l) + Best\_position$\;
            \tcp*{$b$ is constant, $l \in [-1,1]$}
        }
        
        Ensure $Position[i]$ stays within $[LB, UB]$\;
    }
    Update parameter $a \leftarrow 2 - t \cdot (2 / Max\_iters)$\;
}

\BlankLine
\Return{$Best\_position$, $Best\_score$}\;
\end{algorithm}

\section{Experimental analysis}
\label{sec_four}
\subsection{experimental environment}
In order to verify the feasibility of the VSQC-WOA model, this paper conducts a comparative test of the model on the MNIST dataset, the quantum circuit construction and measurement involved are utilized in Paddle-quantum, and the training and testing of the model is done in the Pytorch deep learning framework. The system hardware environment is a 12th Gen Intel(R) core(TM) i5-12600KF processor, NVIDIA GeForce RTX 3060 Ti. The software environment is Python 3.8.18, Pytorch 2.2.0+cu121, Paddle-quantum 2.4.0.

\subsection{Relevant data sets}
The MNIST dataset is a widely used handwritten digit recognition dataset originally created by the National Institute of Standards and Technology to evaluate the performance of handwritten character recognition systems. It contains a large number of handwritten samples from 250 different people, which are divided into a training set, a test set, and a validation set. The MNIST dataset is an entry-level dataset for many machine learning tutorials and research due to its moderate size, ease of handling, and challenging nature.

The MNIST dataset contains 60,000 training samples and 10,000 test samples, each of which is a 28x28 pixel grayscale image. The value of each pixel represents the brightness of that pixel, ranging from 0 (black) to 255 (white). Each image is labeled with a number ranging from 0 to 9. It is a multiclassification problem with 10 different classes.

In the study of this paper, the process of data loading and preprocessing, the training data and test data are extracted from the training and test sets of the MNIST dataset, respectively. 1000 training samples and 200 test samples are selected for each category, whose data are loaded and both training and test samples are randomly disrupted to ensure that the data are consistent and reproducible for each training and testing. The sample images of the MNIST dataset are shown in Figure 8.

\begin{figure}
\centerline{\includegraphics[width=360pt,height=12pc]{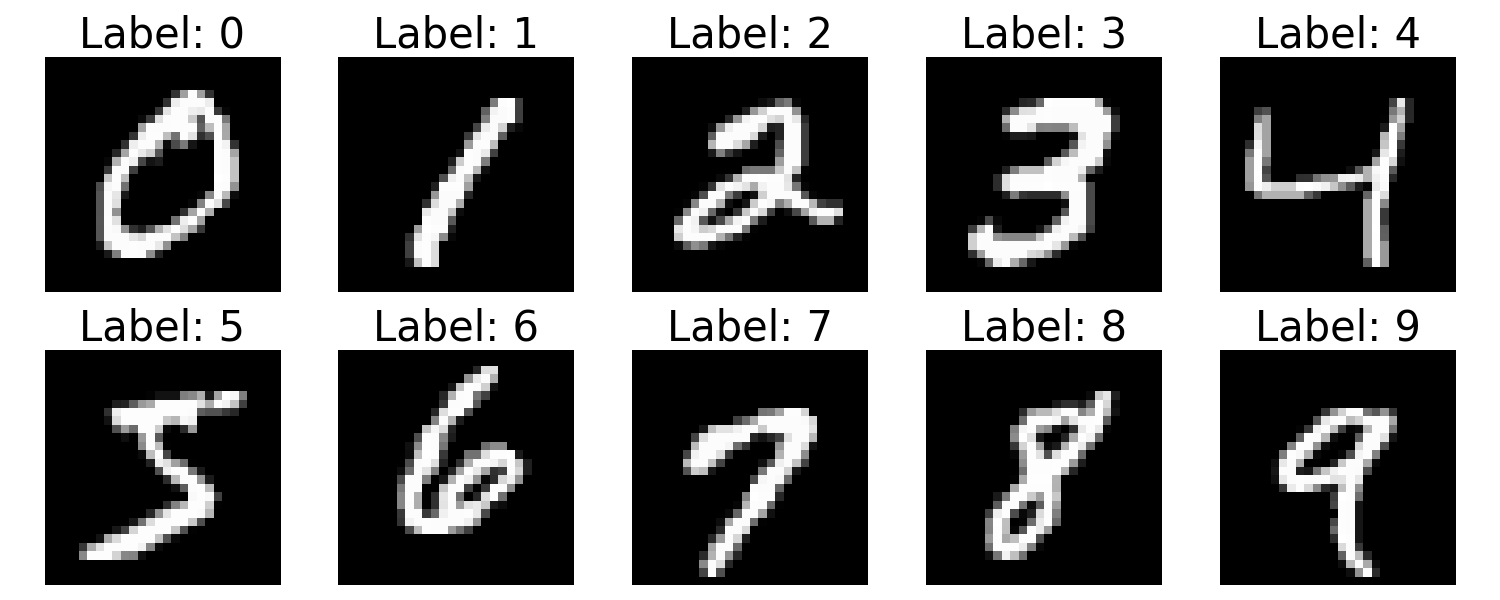}}
\caption{Sample images of the MNIST dataset\label{mnist}}
\end{figure}

\subsection{Setting of hyperparameters}

In this experiment, we carefully set the hyperparameters of the model through several experiments and debugging, aiming to optimize its training efficiency and generalization ability while ensuring its performance. In the process of hyperparameter adjustment, we paid special attention to how to effectively avoid the barren plateau phenomenon as well as the problems of gradient vanishing and gradient explosion, and further improved the stability and expressiveness of the model by reasonably designing the learning rate, initialization parameter distribution and optimizer configuration. All the hyperparameters are shown in Table 1.

\begin{table}
\centering
\caption{\label{tab:table4}Hyper-parameters used for model.  }
 \begin{tabular}{lll}
\hline
Parameter & Values & Remarks\\
\hline
N & 10 & Number of global qubits\\
$n_{qsc}$ & 2 & Width of the shadow circuit\\
D & 3& Depth of circuit\\
Epochs & 20& Number of rounds of training\\
LR & 0.09& Learning rate\\
BATCH & 20 & Training batch size\\
$N_{train}$ & 1000 & Training set size\\
$N_{test}$ & 200 & Testing set size\\
\hline
\end{tabular}
\end{table}

\subsection{Results and analysis}

\subsubsection{\label{sec:levelb}Improved VSQC model for binary classification}

In the experiments in this paper, we randomly selected 10 out of 45 sets of binary classification tasks from the MNIST dataset as the focus of the results presentation. Each group of tasks consists of two types of handwritten digits, aiming to verify the applicability and expressiveness of the model in different binary categorization tasks, as well as to ensure that the experimental results are broadly representative and statistically significant.

In preliminary experiments, we compared the classification effects of five different designs of localized shadow circuits on these 10 sets of binary classification tasks. These shadow circuits are all constructed through unique design concepts to explore the effect of different quantum circuit structures on the classification performance. By comparing the classification accuracies, the results show that Circuit 5 (i.e., the localized shadow circuit we designed) performs the best in all tasks, significantly outperforming the other four shadow circuit structures. The superior performance of Circuit 5 reflects its advantages in feature extraction and information representation.

Based on this result, we further employ Circuit 5 as a core component to construct a shadow circuit structure for variational quantum shadow circuits (VSQCs.) The VSQC model takes full advantage of Circuit 5's capabilities in quantum state manipulation and feature capture, combining the flexibility and scalability of parameterized quantum circuits.

Subsequently, we performed a comprehensive comparison experiment of the Circuit 5-based VSQC model with other existing classical and quantum hybrid models. The experimental results show that the VSQC model exhibits significant advantages in classification performance, not only with higher accuracy, but also demonstrating good generalization ability and stability. This series of experimental results fully verifies the important role of our designed shadow circuit in enhancing the performance of quantum neural networks, and also demonstrates the potential application prospects of quantum computing in image classification tasks. The accuracy results of the ten MNIST datasets are shown in Tables  2 and 3.

\begin{table}
\centering
\caption{\label{tab:table5}Comparison of binary classification accuracy of 5 shadow circuits}
\centering
\centerline{}
 \textbf{}
 \begin{tabular}{cccccc}
\hline
Datasets 
& Circuit-1& Circuit-2& Circuit-3& Circuit-4& Circuit-5
\\
\hline
(0,1)
&     98.5\%& 97.5\%& 99.1\%& 98.9\%& \textbf{100.0\%}
\\
(1,4)
& 98.3\%&\textbf{100.0\%}& \textbf{100.0\%}& 99.4\%& \textbf{100.0\%}
\\
 (1,6)
& 98.3\%& 97.2\%& 99.6\%& 99.9\%&\textbf{100.0\%}
\\
 (1,9)
& 98.1\%& \textbf{100.0\%}& \textbf{100.0\%}& \textbf{100.0\%}&\textbf{100.0\%}
\\
(2,7)
& 86.6\%& 97.1\%& 96.5\%& 98.1\%& \textbf{99.9\%}
\\
(2,8)
& 90.4\%& 94.4\%& 92.3\%& 92.0\%& \textbf{97.4\%}
\\
(4,6)
& 97.9\%& 97.9\%& 99.8\%& 99.4\%& \textbf{100.0\%}
\\
(5,6)
& 94.8\%& 93.2\%& 91.1\%& 95.9\%& \textbf{99.3\%}
\\
(6,8)
& 99.0\%& 98.2\%& \textbf{100.0\%}& 99.2\%& \textbf{100.0\%}
\\
(6,9)& 99.8\%& 98.5\%& 99.4\%& \textbf{100.0\%}& \textbf{100.0\%}
 \\
\hline
\end{tabular}
\end{table}

\begin{table}
\centering
\caption{\label{tab:table6}Comparison of binary classification accuracy between VSQC and other models}
\centering
\centerline{}
 \begin{tabular}{cccccc}
\hline
Datasets & VSQC & CNN & QNN & SVM& RF
\\
\hline
(0,1)&\textbf{100.0\%}& 97.7\%& 96.9\%& 98.0\%& 99.5\%
\\
(1,4)& \textbf{100.0\%}& 97.7\%& 93.3\%& 97.0\%& 98.5\%
\\
 (1,6)& \textbf{100.0\%}& 97.7\%& 92.6\%& 98.5\%&\textbf{100.0\%}
\\
 (1,9)& \textbf{100.0\%}& 97.7\%& 90.6\%& 93.5\%&99.5\%
\\
(2,7)& \textbf{99.9\%}& 96.3\%& 89.2\%& 94.5\%& 96.0\%
\\
(2,8)& \textbf{97.4\%}& 97.1\%& 95.8\%& 90.0\%& 94.0\%
\\
(4,6)& \textbf{100.0\%}& 96.6\%& 92.4\%& 93.5\%& 97.0\%
\\
(5,6)& \textbf{99.3\%}& 97.3\%& 91.8\%& 93.5\%& 95.0\%
\\
(6,8)& \textbf{100.0\%}& 97.0\%& 90.7\%& 94.0\%& 98.5\%
\\
(6,9)& \textbf{100.0\%}& 97.6\%& 94.8\%& 95.5\%& 98.0\%\\
\hline
\end{tabular}
\end{table}

In order to show a more comprehensive results, we conducted simulation experiments on all binary classification tasks with the VSQC model based on Circuit 5, and the results are shown in Figure 9. The experiments show that the classification accuracy of the VSQC model exceeds 95\% on most tasks, and even reaches 100\%
on some tasks, demonstrating excellent classification performance and stability. These results fully validate the key role of the Circuit 5 shadow circuit design in improving the model performance, and also demonstrate the efficiency and adaptability of the VSQC model in handling binary classification tasks.
\begin{figure}
\centerline{\includegraphics[width=440pt,height=10pc]{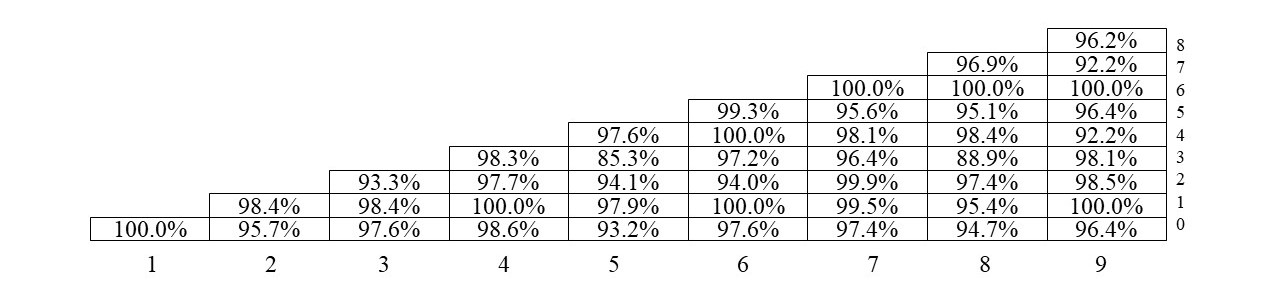}}
\caption{Accuracy of all binary classifications in VSQC based on Circuit 5.\label{figure9}}
\end{figure}

\subsubsection{\label{sec:levelb}Multi-classification of VSQC-WOA models}

In this section, we provide a detailed analysis of the performance of the VSQC-WOA model in multi-categorization tasks, and use the three-categorization task as an experimental example. From the MNIST dataset, we selected four sets of three-categorization tasks, namely $(0,1,2)$,$(1,5,7), (3,4,6),$ and $(6,7,8)$, in order to comprehensively evaluate the model's categorization ability. In the experiments, the same hyperparameter settings are used for all models to ensure fair and comparable results.

Firstly, we conducted comparative experiments on the VSQC-WOA model using a number of different optimization algorithms, including particle swarm optimization algorithm (PSO) [43], genetic algorithm (GAO) [44], artificial immunity algorithm (AIO) [36], and the baseline VSQC model without incorporating WOA. By comparing and analyzing the effects of different optimization strategies on the performance of the VSQC model, we delve into the role of WOA in improving the model performance. The evaluation metrics include classification accuracy, precision, recall, F1 score, and loss value. The comprehensive analysis of these metrics can comprehensively reflect the performance strengths and weaknesses of the VSQC-WOA model in the triple classification task. The experimental results of different algorithms are shown in Figure 10.
\begin{figure}
\centerline{\includegraphics[width=400pt,height=50pc]{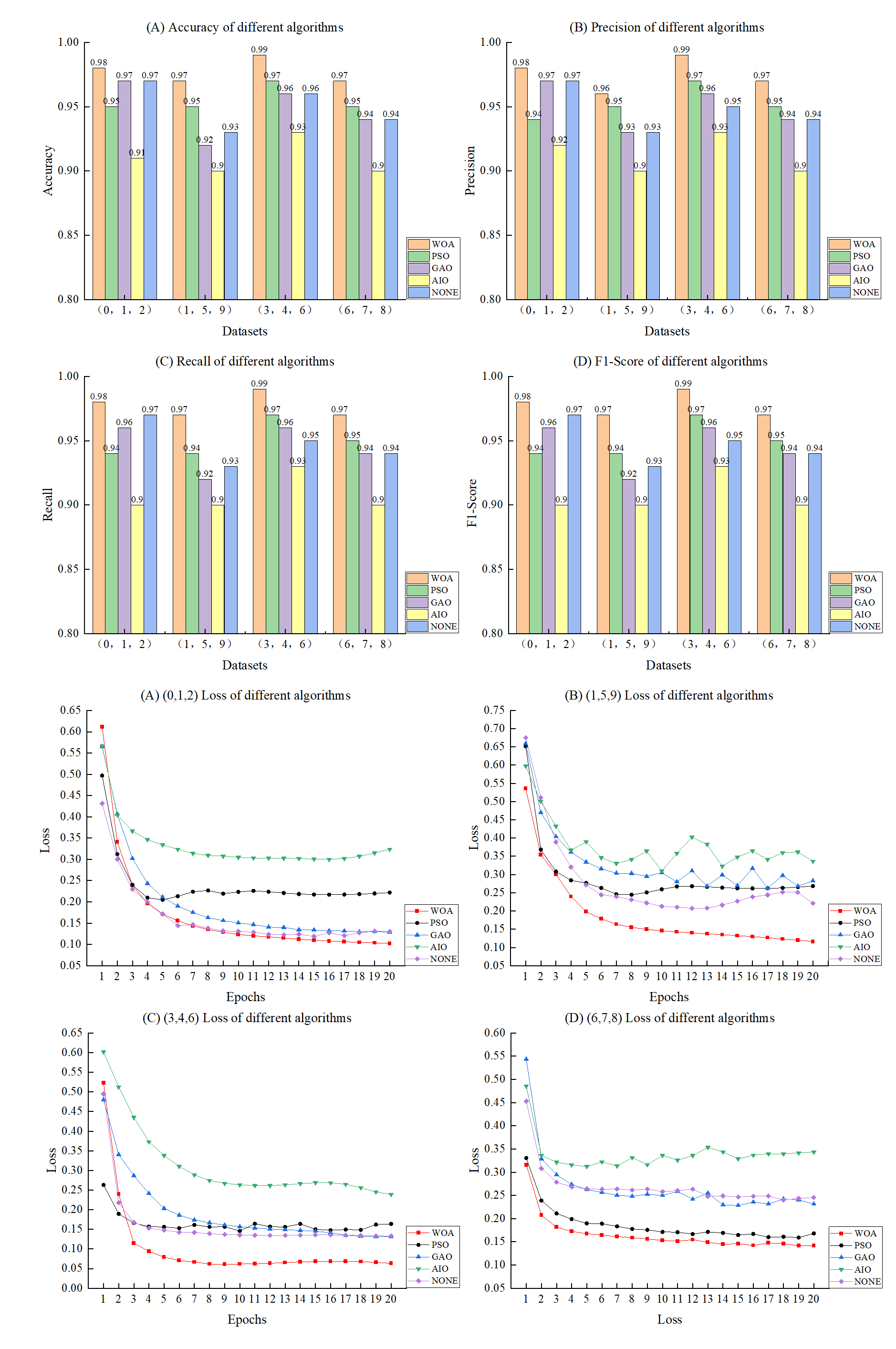}}
\caption{Performance comparison of different algorithms on four sets of datasets.\label{figure=10}}
\end{figure}

From the experimental results, we can see that the WOA-optimized VSQC model performs best in all tasks. In all four metrics, accuracy, precision, recall and F1 score, the WOA optimization model achieves results above 97\%, with some tasks (e.g., $(3,4,6)$ and $(6,7,8)$ reaching the top level of 99\%. In contrast, PSO and GAO are next in circuit, and the performance of AIO and benchmark model (VSQC) is significantly weaker, especially in $(1,5,9)$, where the benchmark model has the lowest performance in each metric, which only reaches about 92\%-94\%. In terms of the trend of the loss values, the WOA-optimized model shows fast convergence ability and stability, with the loss values stabilizing at the lowest level close to 0.05 after about 10 rounds of training. On the other hand, the loss values of PSO and GAO decrease slowly with slight fluctuations in the later stages; the loss values of AIO and the benchmark model VSQC fluctuate more, especially in $(6,7,8)$, which are difficult to converge. These data show that the WOA optimization significantly improves the performance and robustness of the VSQC model, and is the algorithmic scheme with the best optimization effect in the multi-classification task.

In this experiment, we further comprehensively compare the VSQC-WOA model with other classical models, including traditional support vector machines (SVMs), convolutional neural networks (CNNs), quantum neural networks (VQCs), as well as random forests (RF) models and benchmark models (VSQCs), in order to evaluate the overall performance advantages of the VSQC-WOA model. The experimental results of different models are shown in Figure 11.

As can be seen from the figure, the VSQC-WOA model outperforms the other models in all four sets of tasks, and especially shows significant advantages in $(1,5,9)$ and $(6,7,8)$. Specifically, the accuracy of the VSQC-WOA model stays above 98\% in all tasks, while the accuracy of the other models fluctuates considerably, with the accuracy of VQC, CNN, and SVM ranging from 92\%-96\%, while the RF model performs relatively weakly, reaching only 85\%-90\% in some tasks. In the precision and recall metrics, the VSQC-WOA model also performs well, both higher than 97\%, especially in tasks $(3,4,6)$ and $(6,7,8)$ where the precision and recall reach more than 99\% respectively, which is significantly ahead of the other models.The trend of the F1 scores is in circuit with the other metrics, and the performance of the VSQC-WOA model is consistently the highest, which fully verifies its effectiveness and superiority in the multiple classification task, fully verifying its effectiveness and superiority. These results show that the VSQC-WOA model is not only significantly competitive in classification performance, but also outperforms the traditional classical model and other quantum models in terms of generalization ability and stability.

\begin{figure}
\centerline{\includegraphics[width=450pt,height=23pc]{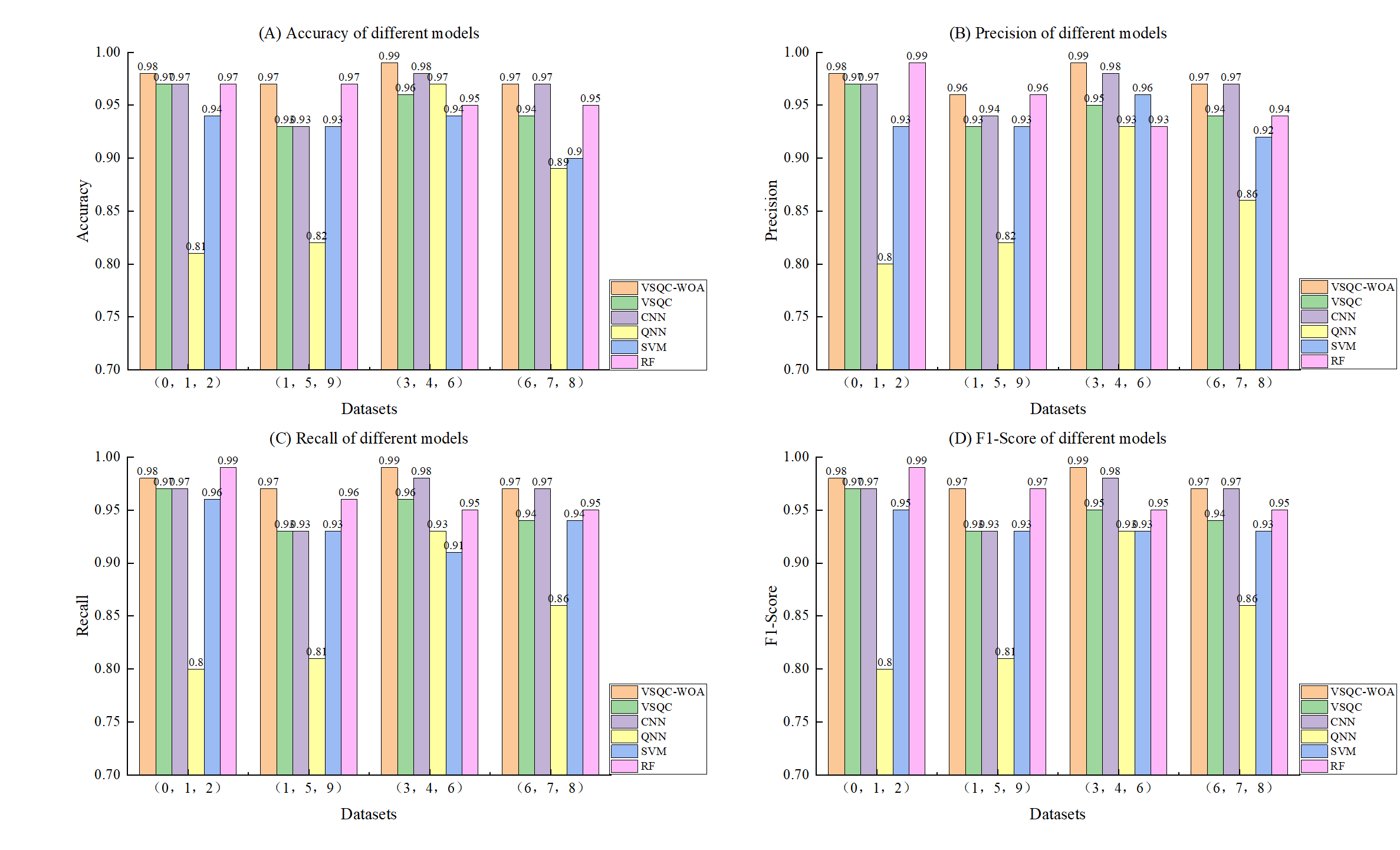}}
\caption{Performance comparison of different models on four sets of datasets.\label{figure=11}}
\end{figure}

In conclusion, through the detailed comparisons under four sets of three classification tasks demonstrated by multiple comparison experiments, we further validate the effectiveness of WOA optimization in enhancing the performance of quantum models, provide strong support for the application of the VSQC-WOA model to multi-classification problems, and also show that the model has a wide range of application potentials.

\section{Conclude}

In this paper, a variational shadow quantum circuit (VSQC-WOA) model based on the whale optimization algorithm is proposed with the aim of improving the performance of quantum neural networks in image classification tasks. By designing strongly entangled variational quantum circuits combined with local shadow feature extraction and sliding mechanism, the model is able to efficiently characterize global features and provide rich quantum feature representation for classification tasks. In addition, the Whale Optimization Algorithm (WOA) is introduced to further optimize the weights and biases of the classical fully-connected neural network, which significantly improves the model's representational ability and classification accuracy.

Experimental results show that the VSQC-WOA model performs well in both binary and multi-classification tasks on the MNIST dataset. In the binary classification task, the classification accuracy of the VSQC-WOA model exceeds 95\% on most tasks, and even reaches 100\% on some tasks. In the multi-classification task, the VSQC-WOA model outperforms other classical and quantum models in terms of accuracy, precision, recall and F1 score, demonstrating excellent classification performance and robustness.

The innovations in this paper include the design of strongly entangled quantum state evolution circuits, the optimization of classical post-processing network parameters by combining with the whale optimization algorithm, and the verification of the superiority of the WOA algorithm in image classification tasks through comparative experiments. These innovations not only enhance the performance of quantum neural networks, but also provide new ideas for the integration of quantum computing and classical machine learning.

Although the VSQC-WOA model performed well in the experiment, there are still some aspects that need to be improved. Firstly, the high cost of quantum computing resources creates a limitation on the application of the model to large-scale datasets and complex tasks, while the parameter tuning process of the model is complicated and needs to be further simplified and automated. Secondly, the generalization ability of the model on other datasets and tasks needs to be further verified and improved. In addition, how to effectively combine the advantages of quantum and classical computing in order to optimize the design and implementation of hybrid architectures is also a key issue that needs to be addressed.

Future research can further optimize and extend the VSQC-WOA model in the following directions: firstly, to improve the robustness and adaptability of the model in noisy environments to cope with the limitations of quantum hardware; second, to extend the scope of the model's application, and to test its performance in larger and more complex datasets (e.g., CIFAR-10 or ImageNet); third, to explore more efficient optimization algorithms or to combine multiple optimization strategies to improve the performance; fourth, to study the theoretical foundation of quantum neural networks in depth, combining quantum information theory and classical machine learning theory to enhance the theoretical interpretability of the model and support its further development.

\section*{Statement}
\textbf{Acknowledgments}

This work was supported by National Natural Science Foundation of China (No.61772295), Key Projects of Chongqing Natural Science Foundation Innovation Development Joint Fund (CSTB2023NSCQ-LZX0139), Open Fund of Advanced Cryptography and System Security Key Laboratory of Sichuan Province (Grant No. SKLACSS--202208) and Technology Research Program of Chongqing Municipal Education Commission (Grant no. KJZD-M202000501).

\textbf{Author contributions statement}
Shuang Wu proposed ideas for the paper and wrote a manuscript. Xueliang Song translated and formatted the paper, Yumin Dong provided necessary financial support, and Fanghua Jia created images and verified experimental data.

\textbf{Competing interests statement}
The author(s) declare no competing interests.

\textbf{Data availability statement}
The datasets generated during and/or analyzed during the current study are available from the corresponding author on reasonable request.
\bibliography{main}

\begin{thebibliography}{99}\footnotesize
\itemsep=-3pt plus.2pt minus.2pt

\bibitem{1} LeCun Y, Bengio Y, Hinton G. Deep learning. \href{https://www.nature.com/articles/nature14539}{2015 
\emph{nature, 521(7553): 436-444.}}

\bibitem{2} Biamonte J, Wittek P, Pancotti N, et al. Quantum machine learning.  
\href{https://www.nature.com/articles/nature23474}{2017 \emph{Nature,549(7671): 195-202.}}

\bibitem{3} Cerezo M, Verdon G, Huang H Y, et al. Challenges and opportunities in quantum machine learning. \href{https://www.nature.com/articles/s43588-022-00311-3}{2022 \emph{Nature Computational Science, 2(9): 567-576.}}

\bibitem{4} Jerbi S, Gyurik C, Marshall S C, et al. Shadows of quantum machine learning.
\href{https://www.nature.com/articles/s41467-024-49877-8}{2024 \emph{Nature Communications, 15(1): 5676.}}

\bibitem{5} Pulicharla M R. Hybrid Quantum-Classical Machine Learning Models: Powering the Future of AI.
\href{https://doi.org/10.55662/JST.2023.4102}{2023 
\emph{Journal of Science \& Technology, 4(1): 40-65.}}

\bibitem{6} Wang A, Hu J, Zhang S, et al. Shallow hybrid quantum-classical convolutional neural network model for image classification.   
\href{https://doi.org/10.1103/PhysRevApplied.14.064020}{2024 
\emph{Quantum Information Processing, 23(1): 17.}}

\bibitem{7} Xiang Q, Li D, Hu Z, et al. Quantum classical hybrid convolutional neural networks for breast cancer diagnosis.
\href{https://doi.org/10.1038/s41598-024-74778-7}{2024 
\emph{Scientific Reports, 14(1): 24699.}}

\bibitem{8} Shin M, Lee J, Jeong K. Estimating quantum mutual information through a quantum neural network.     \href{https://link.springer.com/article/10.1007/s11128-023-04253-1}{2024 \emph{Quantum Information Processing, 23(2): 57.}}

\bibitem{9} Zhou M G, Liu Z P, Yin H L, et al. Quantum neural network for quantum neural computing.    \href{https://spj.science.org/doi/full/10.34133/research.0134}{2023 
\emph{Research, 6: 013}}

\bibitem{10}Li W, Lu Z, Deng D L. Quantum neural network classifiers: A tutorial.    
\href{https://doi.org/10.21468/SciPostPhysLectNotes.61}{2022 
\emph{SciPost Physics Lecture Notes: 061.}}

\bibitem{11} Kapoor A, Wiebe N, Svore K. Quantum perceptron models.    \href{https://proceedings.neurips.cc/paper/2016/hash/d47268e9db2e9aa3827bba3afb7ff94a-Abstract.html}{2019
\emph{Advances in neural information processing systems, 29.}}

\bibitem{12} Schütt K T, Arbabzadah F, Chmiela S, et al. Quantum-chemical insights from deep tensor neural networks. \href{https://www.nature.com/articles/ncomms13890}{2017 
\emph{ Nature communications, 8(1): 13890.}}

\bibitem{13} Yan S, Qi H, Cui W. Nonlinear quantum neuron: A fundamental building block for quantum neural networks.    \href{https://doi.org/10.1103/PhysRevA.102.052421}{2020 
\emph{Physical Review A, 102(5): 052421.}}

\bibitem{14} Farhi E, Neven H. Classification with quantum neural networks on near term processors.     
\href{https://arxiv.org/abs/1802.06002}{2018 
\emph{arXiv preprint arXiv:1802.06002.}}

\bibitem{15} Grant E, Benedetti M, Cao S, et al. Hierarchical quantum classifiers.   
\href{https://www.nature.com/articles/s41534-018-0116-9}{2018 
\emph{npj Quantum Information, 4(1): 65.}}

\bibitem{16} Cong I, Choi S, Lukin M D. Quantum convolutional neural networks.   
\href{https://www.nature.com/articles/s41567-019-0648-8}{2019 
\emph{Nature Physics, 15(12): 1273-1278.}}

\bibitem{17} Li Y C, Zhou R G, Xu R Q, et al. A quantum deep convolutional neural network for image recognition. 
\href{https://iopscience.iop.org/article/10.1088/2058-9565/ab9f93/meta}{2020 
\emph{Quantum Science and Technology, 5(4): 044003.}}

\bibitem{18} Henderson M, Shakya S, Pradhan S, et al. Quanvolutional neural networks: powering image recognition with quantum circuits.
\href{https://link.springer.com/article/10.1007/s42484-021-00056-8}{2020 
\emph{Quantum Machine Intelligence, 2(1): 2.}}

\bibitem{19} Kashif M, Al-Kuwari S. Design Space Exploration of Hybrid Quantum–Classical Neural Networks. 
\href{https://www.mdpi.com/2079-9292/10/23/2980}{2021 
\emph{Electronics, 10(23): 2980.}}

\bibitem{20} Pesah A, Cerezo M, Wang S, et al. Absence of barren plateaus in quantum convolutional neural networks.  \href{https://journals.aps.org/prx/abstract/10.1103/PhysRevX.11.041011}{2021
\emph{Physical Review X, 11(4): 041011.}}

\bibitem{21} Abbas A, Sutter D, Zoufal C, et al. The power of quantum neural networks.   
\href{https://www.nature.com/articles/s43588-021-00084-1}{2021 
\emph{Nature Computational Science, 1(6): 403-409.}}

\bibitem{22} Qu Z, Shi W, Liu B, et al. IoMT-based smart healthcare detection system driven by quantum blockchain and quantum neural network.  \href{https://ieeexplore.ieee.org/document/10172019}{2023 
\emph{IEEE journal of biomedical and health informatics.}}

\bibitem{23} Song Z, Xu J, Zhou X, et al. Transforming two-dimensional tensor networks into quantum circuits for supervised learning.  
\href{https://iopscience.iop.org/article/10.1088/2632-2153/ad2fec/meta}
{2024 
\emph{Machine Learning: Science and Technology, 5(1): 015048.}}

\bibitem{24} Park S, Baek H, Yoon J W, et al. AQUA: Analytics-driven quantum neural network (QNN) user assistance for software validation.   
\href{https://doi.org/10.1016/j.future.2024.05.047}{2024 
\emph{Future Generation Computer Systems.}}

\bibitem{25}Fan F, Shi Y, Guggemos T, et al. Hybrid quantum-classical convolutional neural network model for image classification.    
\href{https://ieeexplore.ieee.org/abstract/document/10254235}{2023 
\emph{IEEE transactions on neural networks and learning systems.}}

\bibitem{26}Wu Q, Liu W, Huang Y, et al. A degressive quantum convolutional neural network for quantum state classification and code recognition.  
\href{https://doi.org/10.1016/j.isci.2024.109394}{2024 
\emph{Iscience, 27(4).}}

\bibitem{27} Shi M, Situ H, Zhang C. Hybrid quantum neural network structures for image multi-classification.   
\href{https://iopscience.iop.org/article/10.1088/1402-4896/ad3e3d/meta}{2024 
\emph{Physica Scripta, 99(5): 056012.}}

\bibitem{28} Wu Y, Feng J. Development and application of artificial neural network.  
\href{https://doi.org/10.1007/s11277-017-5224-x}{2018 
\emph{Wireless Personal Communications, 102: 1645-1656.}}

\bibitem{29} Yegnanarayana B. Artificial neural networks.  
\href{https://www.google.com/books/edition/Artificial_Neural_Network_for_Drug_Desig/_8_UBQAAQBAJ?kptab=editions&sa=X&ved=2ahUKEwiwrJef042LAxXMIUQIHUNBJrsQmBZ6BAgEEAs&cshid=1737697671667490}{2009 
\emph{PHI Learning Pvt. Ltd.}}

\bibitem{30} Schuld M, Sinayskiy I, Petruccione F. The quest for a quantum neural network.   
\href{https://doi.org/10.1007/s11128-014-0809-8}{2014 
\emph{Quantum Information Processing, 13: 2567-2586.}}

\bibitem{31} Dong Y, Wu S. Hybrid quantum neural network based on weight remapping and its applications.   
\href{https://doi.org/10.1088/1402-4896/ad9ae1}{2024 
\emph{Physica Scripta, 100(1): 015114.}}

\bibitem{32} Steane A. Quantum computing  
\href{https://iopscience.iop.org/article/10.1088/0034-4885/61/2/002/meta}{1998 
\emph{Reports on Progress in Physics, 61(2): 117.}}

\bibitem{33} DiVincenzo D P. Quantum gates and circuits.  
\href{https://doi.org/10.1098/rspa.1998.0159}{1998 
\emph{Series A: Mathematical, Physical and Engineering Sciences, 454(1969): 261-276.}}

\bibitem{34} Li, Guangxi, Zhixin Song, and Xin Wang. "VSQL: Variational Shadow Quantum Learning for Classification."    
\href{https://doi.org/10.1609/aaai.v35i9.17016}{2021 
\emph{Proceedings of the AAAI conference on artificial intelligence. 35(9): 8357-8365.}}

\bibitem{35} Jerbi S, Gyurik C, Marshall S C, et al. Shadows of quantum machine learning.    
\href{https://doi.org/10.1038/s41467-024-49877-8}{2024 
\emph{Nature Communications, 15(1): 5676.}}

\bibitem{36} Dong Y, Zhu T, Fu Y, et al.Variational shadow quantum neural network based on immune optimisation algorithm.     
\href{https://doi.org/10.1007/s11128-024-04363-4}{2024 
\emph{Quantum Information Processing,  23(5): 155.}}

\bibitem{37} Mirjalili S, Lewis A. The whale optimization algorithm.      
\href{https://doi.org/10.1016/j.advengsoft.2016.01.008}{2016 
\emph{Advances in engineering software, 95: 51-67.}}

\bibitem{38} Aljarah I, Faris H, Mirjalili S. Optimizing connection weights in neural networks using the whale optimization algorithm.       
\href{https://doi.org/10.1007/s00500-016-2442-1}{2018 
\emph{Soft Computing, 22: 1-15.}}

\bibitem{39} Li G, Zhao X, Wang X. Quantum self-attention neural networks for text classification.      
\href{https://doi.org/10.1007/s11432-023-3879-7}{2024 
\emph{Science China Information Sciences, 67(4): 142501.}}

\bibitem{40} Schuld M, Bocharov A, Svore K M, et al. Circuit-centric quantum classifiers.      
\href{https://doi.org/10.1103/PhysRevA.101.032308}{2020 
\emph{Physical Review A, 101(3): 032308.}}

\bibitem{41} Mitarai K, Negoro M, Kitagawa M, et al. Quantum circuit learning.      
\href{https://doi.org/10.1103/PhysRevA.98.032309}{2018 
\emph{Physical Review A, 98(3): 032309.}}

\bibitem{42} Wierichs D, Izaac J, Wang C, et al. General parameter-shift rules for quantum gradients.      
\href{https://doi.org/10.22331/q-2022-03-30-677}{2022 
\emph{Quantum, 6: 677.}}

\bibitem{43} Wang D, Tan D, Liu L. Particle swarm optimization algorithm: an overview.      
\href{https://doi.org/10.1007/s00500-016-2474-6}{2018 
\emph{Soft computing, 22(2): 387-408.}}

\bibitem{44} Mirjalili S. Genetic algorithm.     
\href{https://doi.org/10.1007/s11432-023-3879-7}{2019 
\emph{Evolutionary algorithms and neural networks: theory and applications, 43-55.}}

\end{thebibliography}

\end{document}